\documentclass[aps, 
pra,
twocolumn,
longbibliography
]{revtex4-1}
\usepackage{graphicx}
\usepackage{amsmath}
\usepackage{amssymb,bm}
\usepackage{color}
\usepackage{hyperref}
\hypersetup{colorlinks,linkcolor={blue},citecolor={red},urlcolor={blue}}
\usepackage{tikz}
\usepackage[caption=false]{subfig}

\def\O{\mathcal{O}}

\def\ave#1{\left\langle #1 \right\rangle}

\def\O{\mathcal{O}}

\newcommand{\brac}[1]{\left\lbrace #1
	\right\rbrace}

\begin{document}

\title{Linear multiport photonic interferometers: loss analysis of temporally-encoded architectures}

\author{Haoyu Qi}
\author{Lukas G. Helt}
\author{Daiqin Su}
\author{Zachary Vernon}
\author{Kamil Br\'adler}
%\author{Daiqin Su}
%\email{daiqin@xanadu.ai}
%\author{Christian Weedbrook}
%\email{christian@xanadu.ai}
%\author{Kamil Br\'adler}
%\email{kamil@xanadu.ai}
\affiliation{Xanadu, 372 Richmond Street West, Toronto, Ontario M5V 1X6, Canada }

%\pacs{03.70.+k, 03.65.Ud, 04.62.+v}

%\date{\today}

\begin{abstract} 
Implementing spatially-encoded universal linear multiport interferometers on integrated photonic platforms with high controllability becomes increasingly difficult as the number of modes increases. In this work, we consider an architecture in which temporally-encoded modes are injected into a chain of reconfigurable beamsplitters and delay loops, and provide a detailed loss analysis for this and other comparable architectures. This analysis also yields a straightforward and explicit description of the control sequence needed for implementing an arbitrary temporally-encoded transformation. We consider possible physical implementations of these architectures and compare their performance. We conclude that implementing a chain of reconfigurable beamsplitters and delay loops on an integrated lithum niobate platform could outperform comparable architectures in the near future.
\end{abstract}

\maketitle

\section{Introduction}
Photonic linear multiport interferometers play a crucial role in metrology, telecommunications, and fundamental science. While often imagined and realized as acting on spatial modes, it is also possible to consider them acting on any degree of freedom of the radiation field. In particular, there has been recent interest in linear multiport interferometers that act on \textit{time-bins}~\cite{motes2014scalable,motes2015implementing,rohde2015simple,he2017time,takeda2017universal}, the temporal degree of freedom of the electromagnetic field.

\begin{figure}[!htbp]
\centering
\includegraphics[scale=0.35]{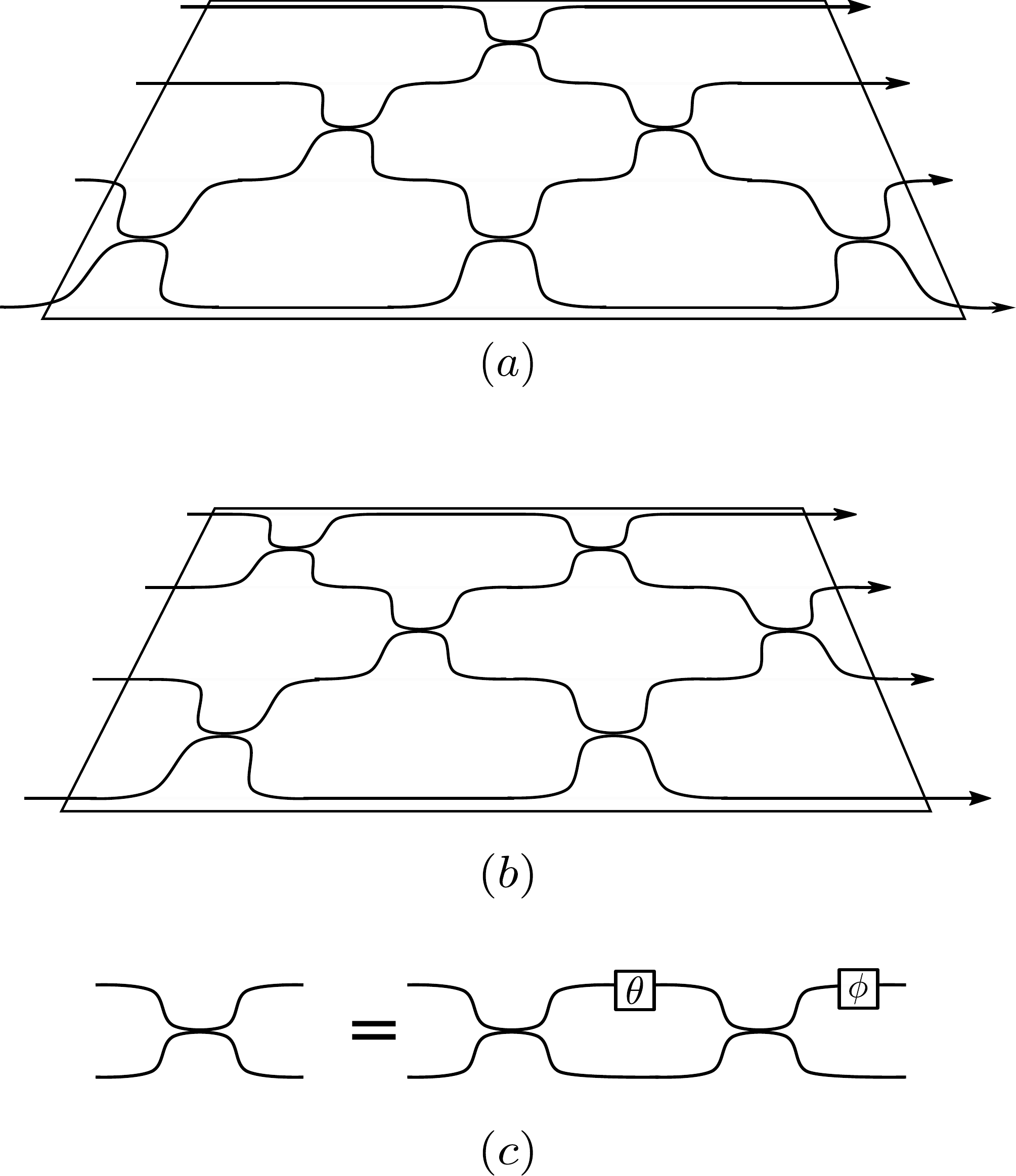}
\caption{Spatially encoded decompositions of $U(4)$. (a)~Unitary decomposed according to Reck's scheme~\cite{reck1994experimental}. (b)~Unitary decomposed according to Clements' scheme\cite{clements2016optimal}.  (c)~Each $SU(2)$ transformation is implemented via a MZI with two controllable phases. \label{fig:spatial-encoding}}
\end{figure}

The standard way to implement an optical unitary transformation acting on a large number of modes, $U(N)$ for large $N$, is to decompose it into a series of $SU(2)$ and $ U(1)$ unitary transformations. It was shown by Reck, et al.~\cite{reck1994experimental} that a universal $N$-mode interferometer can be built out of $N(N-1)/2$ beamsplitters and $N$ phase shifters (see Fig.~\ref{fig:spatial-encoding} (a)). The operation of the phase shifters is often considered separate from the operation of the multiport interferometer, and so we make this distinction here as well, focusing on the $SU(2)$ transformations. For reconfigurability and control, in practice these beamsplitters are commonly implemented as Mach-Zehnder interferometers (MZIs) with two controllable phases  (see Fig.~\ref{fig:spatial-encoding} (c)). More recently, Clements, et al.~\cite{clements2016optimal} proposed another way of decomposing $U(N)$ into $N(N-1)/2$ beamsplitters and $N$ phase shifters enabling a more compact spatial implementation (see Fig.~\ref{fig:spatial-encoding} (b)); other decomposition methods have been proposed as well~\cite{de2018simple}. However, the engineering and control of large spatially encoded multiports is nontrivial, even if implemented on a photonic chip~\cite{harris2016large,harris2017quantum}, and, as such, alternative architectures have also been explored.

\begin{figure*}[htbp!]
	\subfloat[]{%
		\includegraphics[scale=0.7]{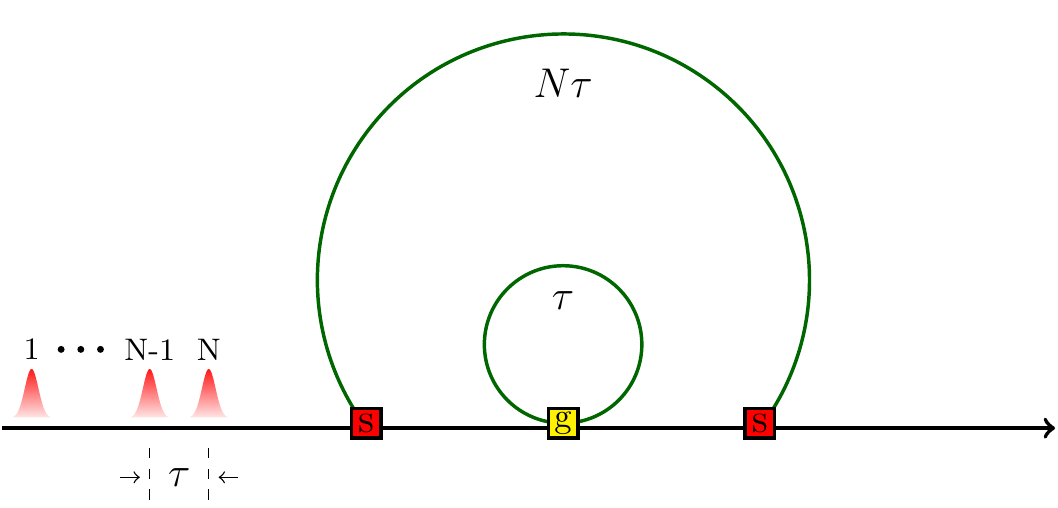}%
	}\hspace{1cm}
	\subfloat[]{%
		\includegraphics[scale=0.6]{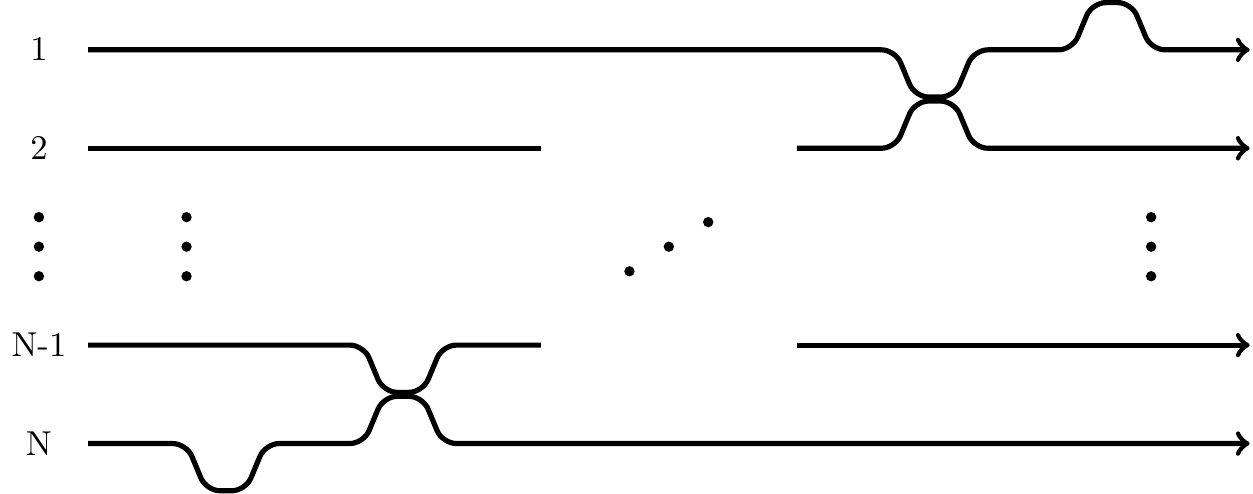}%
	}
	\caption{(a) A schematic of the dual-loop architecture. The yellow box labelled with ``g" indicates a fast tunable MZI and the red boxes labelled with ``s" fast tunable switches. (b) One layer of interactions implemented by the inner loop. Except interactions between all adjacent pulses, the first and the last pulse needs to be pushed into and out from the inner loop. Effectively, the gate plays the role of switches at two ends, which is shown by two ``half'' gates. \label{fig:dual-loop}}
\end{figure*}
One alternative architecture to the spatial encodings mentioned above is the so-called \textit{dual-loop} architecture proposed for Boson Sampling~\cite{motes2014scalable,motes2015implementing}, and later experimentally implemented~\cite{he2017time}. It makes use of a temporal encoding, and requires just a single fast reconfigurable beamsplitter. A scheme to achieve universal quantum computation based on the dual-loop architecture and measurement-based non-Gaussian gates was also proposed~\cite{rohde2015simple,takeda2017universal}. As a second alternative, in this work we consider a \textit{chain-loop} architecture, requiring a sequence of $N-1$ fast reconfigurable beamsplitters, but potentially incurring lower loss and being more amenable to integration on a photonic chip than the dual-loop architecture.

%The rest of this paper is organized as follows. 
In Sec.~\ref{sec:architectures} we review the dual-loop architecture and introduce the chain-loop architecture in detail. We consider all possible losses intrinsic to each architecture and define intuitive expressions to quantify their overall loss, enabling straightforward and direct comparisons. In Sec.~\ref{sec:simulation}, using both analytic and numerical tools, we argue that these expressions can be understood as useful heuristic approximations to the total loss averaged over all possible optical paths.
In Sec.~\ref{sec:implementations}, we consider free-space and chip-based implementations of each loop architecture, generating sets of possible numbers for use in formulae developed in the two previous Sections. Finally, we briefly discuss possible applications of our chain-loop architecture  in Sec.~\ref{sec:application} before concluding in Sec.~\ref{sec:conclusions}.

\section{Loop-based architectures}\label{sec:architectures}
In this section, we consider two possible architectures that make use of time-bin encoding to implement a multiport interferometer. We first briefly review a spatially-encoded architecture where the loss analysis is particularly simple. For each loop-based architecture, we develop a method to show clearly how the loss accumulates compared to the spatial architecture. We define  expressions to quantify the overall transmission of each architecture based on physical intuition, allowing one to easily make comparisons.

\subsection{Spatially-encoded architecture}
Although we focus on loop-based architectures in this work, it is still worthwhile to analyze the loss in a spatially-encoded architecture. On the one hand, the performance of spatial encoding on integrated photonic platforms provides a benchmark for loop-based architectures. On the other hand, it builds intuition for the more complicated loss analysis of the loop-based architectures. Later we will see that we can understand the effect of loss in loop-based architectures as simply related to that in a spatially-encoded architecture. 

For an integrated platform implementation, the main source of loss is often propagation loss. This loss is proportional to the length of an optical path. We note that there can be additional losses at each MZI due to bending and cross-talk, however we ignore them here since, in principle, MZIs can be made propagation-loss-limited. It can be difficult to define a single expression to quantify the overall loss of this architecture, as light can travel along many different paths through the chip. However we find that a useful choice is to use the loss suffered in traveling across the entire chip. Using this definition of effective loss, we write the overall transmission as
\begin{align}\label{eq:transmission-spatial-encoding}
\eta_{SE} = \exp{\left(-\alpha_{C}L_{\text{MZI}}\right)}^N = \eta_{g}^N, 
\end{align}
where $\alpha_{C}$ is the loss coefficient per-unit length, depending on the material from which the interferometer is fabricated, and $L_{\text{MZI}}$ is the length of each MZI. From Fig.~\ref{fig:spatial-encoding} we can see that, theoretically, the minimum chip size is equal to $NL_\text{MZI}$ while, in practice, the chip size is usually much larger due to the electronic inputs for each MZI~\cite{harris2016large}. However, this additional size is not a fundamental limitation, and the purpose of this Subsection is simply to provide a reasonable upper bound for overall architecture transmission,  so we do not consider a larger chip size here. We note that a second intuitive way to understand Eq.~\eqref{eq:transmission-spatial-encoding} is to divide the chip into $N$ layers of gates (here each gate is one MZI), giving $\eta_{g}=\exp{\left(-\alpha_{C}L_{\text{MZI}}\right)}$ the physical meaning of transmission per layer.
\begin{figure*}[!htbp]
	\includegraphics[scale=0.45]{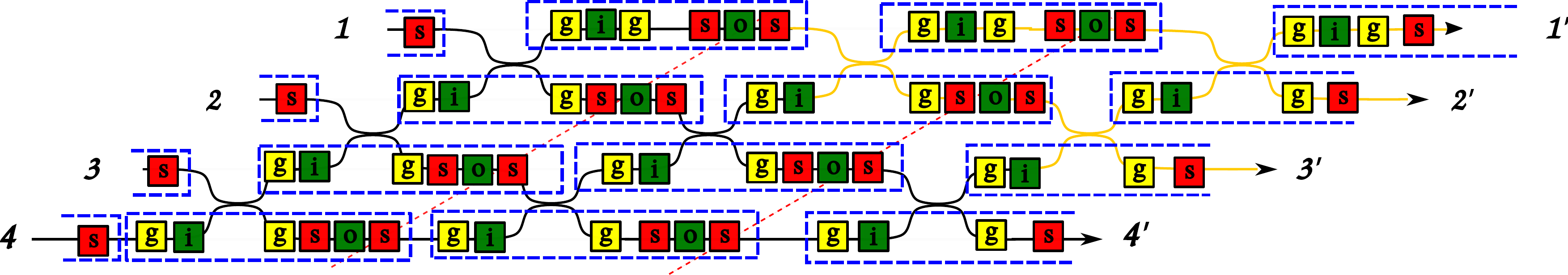}
	\caption{ Implementation of a 4-mode interferometer with a dual-loop architecture. Each time all 4 time-bins travel through the inner loop, one layer of the Reck's scheme \cite{reck1994experimental} is applied. Different layers are separated by red dashed lines. Losses resulting from different optical components are indicated by different boxes: red boxes labeled by an ``s'' for switches, yellow boxes labeled by a ``g'' for MZIs, and green boxes labeled by ``i'' or ``o'' for inner and outer loops, respectively. A detailed procedure outlining how to construct this figure and others like it is given in the main text.}
	\label{fig:double-loop-loss}
\end{figure*}
\subsection{Dual-loop architecture}\label{subsec:double-loop}
The dual-loop architecture, proposed by Motes, et al.~\cite{motes2014scalable,motes2015implementing}, consists of a single fast tunable beamsplitter or MZI, two fast switches, one inner delay loop, and one (larger) outer delay loop (see Fig.~\ref{fig:dual-loop}~(a)). All interactions between different temporal modes happen at the tunable beamsplitter (yellow box in Fig.~\ref{fig:dual-loop}~(a)). Taking the input time-bins to be separated by time $\tau$, the inner loop must implement a delay of duration $\tau$. A single pass of temporal modes through the inner loop implements the equivalent of a layer of beamsplitters of the Reck scheme \cite{reck1994experimental} of Fig.~\ref{fig:spatial-encoding}~(b) (see Fig.~\ref{fig:dual-loop}~(b)). Switches (red boxes in Fig.~\ref{fig:dual-loop}~(a)) and the outer loop guide time-bins back to the inner loop to implement more layers, an operation that must be performed $N-1$ times to realize a universal $N$-mode linear interferometer. As the outer loop must be able to contain all $N$ time-bins, it needs to implement a delay at least $N$ times longer than the inner loop (i.e., $N\tau$). 

From Fig.~\ref{fig:dual-loop}~(a) we see that losses in the dual-loop architecture can occur in any of four different components. We use $\eta_{g}$, $\eta_{i}$, $\eta_{o}$, and $\eta_{s}$ to denote the transmission or efficiency of the MZI (tunable $SU(2)$ gate), the inner loop, the outer loop, and the switches, respectively. Before a time-bin can interfere with the one behind it, it must pass through a switch, the MZI, and the inner loop. Then it can interfere with the subsequent time-bin at the MZI, before passing through a second switch. To implement another layer, the time-bin must pass through the outer loop, after which it is ready to repeat the process of first switch, MZI, and inner loop. 

To enable a more direct comparison with the Reck scheme, and make the details clear, we develop a procedure for inserting losses associated with the dual-loop architecture directly on top of a sketch of a Reck scheme (see Fig.~\ref{fig:double-loop-loss} for the 4-mode case). In principle, time-bin $j$ can be released by the second switch at the $j^\text{th}$ layer. However, doing so causes the time-bins to lose their uniform spacing in time, requiring more complicated detection schemes. We note that not allowing the time-bins to be released ``early'' effectively increases the number of elements some time-bins must pass through (shown in yellow in Fig.~\ref{fig:double-loop-loss}), which in turn introduces more loss, but that it does not change the optical depth, or lossiest path, of the architecture. Additionally, more balanced losses can be useful for some applications~\cite{clements2016optimal}. For all of these reasons, we defer the analysis of architectures that switch time-bins out to detectors as early as possible to future work.

Using red boxes labeled by an ``s'' to denote switches, yellow boxes labeled by a ``g'' to denote MZIs, and green boxes labeled by an ``i'' or an ``o'' to denote inner and outer loops, respectively, the procedure to mock up a Reck scheme to include dual-loop architecture losses is as follows:
\begin{enumerate}
	\item Sketch the corresponding Reck decomposition for $U(N)$. 
	\item Add dashed diagonal lines (shown in red in  Fig.~\ref{fig:double-loop-loss}) to separate the $N-1$ layers. 
        	\item Add $j-1$ MZIs (shown in yellow in  Fig.~\ref{fig:double-loop-loss}) to each layer $j$ so that each layer contains $N-1$ MZIs. 
	\item Add a gate loss to each of the two output modes of each MZI ($2\left(N-1\right)^{2}$ in total).
	\item At each intersection between a dashed line and quantum wire, add two switch losses and one outer loop loss ($N\left(N-2\right)$ sets in total). 
	\item Within each layer, add an inner loop loss to the lower input mode and/or upper output mode of each MZI ($N^{2}$ in total).
	\item Within each layer, add just two gate losses: one on the lower input mode to the lowest MZI, and one on the upper output mode or the uppermost MZI. These two extra losses are needed to push the first time-bin into the inner loop, and to push the last time-bin out of the inner loop.
	\item Finally, add a switch loss to each mode at the beginning and end of the sketch. These indicate that each time-bin needs to pass through a switch whenever it enters or exists the architecture. 
\end{enumerate}
Note that one can also begin with the Clements' decomposition and, following a similar procedure, end up with the same diagram. 

Similar to the definition of Eq.~\eqref{eq:transmission-spatial-encoding}, we define the effective overall transmission of the chain-loop architecture to be that experienced in travelling the entire interferometer from input mode $i$ to output mode $i^{\prime}$ (see Fig.~\ref{fig:double-loop-loss}). As can be seen in  Fig.~\ref{fig:double-loop-loss}, it has the nice feature that it is the same for each mode and independent of the path taken from input to output. In particular, for an $N=4$ interferometer, each time-bin passes through 6 MZIs, 6 switches, 3 inner loops, and 2 outer loops. In general, this can be written as
\begin{equation}\label{eq:double-loop-overall-loss}
\eta_{\text{DL}} = \left(\eta_{g}^2\eta_{s}^2\eta_{i}\eta_{o}\right)^{N-1}\eta_{o}^{-1}.
\end{equation}  

In Sec.~\ref{sec:implementations} we estimate the size of these transmission factors by considering possible realizations in the laboratory. As compared to spatial encodings, the required number of physical beamsplitters in the dual-loop architecture is greatly reduced. However, the introduction of delay loops, in particular the outer loop, as well as fast reconfigurable components, may increase photon loss.

\subsection{Chain-loop architecture}\label{subsec:chain-loop}
\begin{figure}[h]
	\includegraphics[width=8.5 cm]{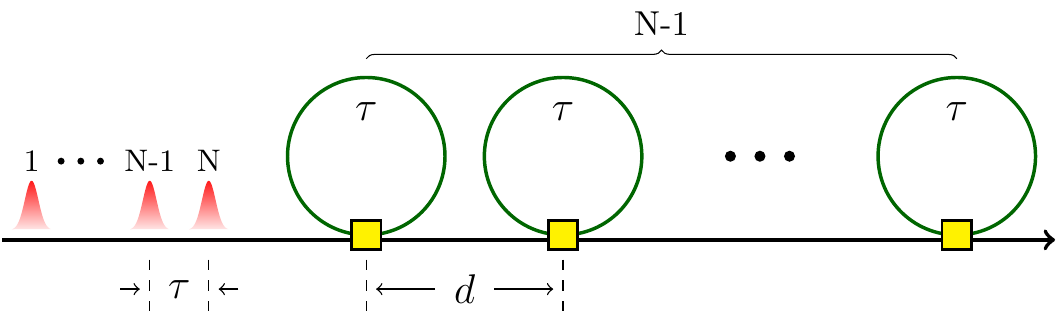}
	\caption{ A schematic of the chain-loop architecture. The yellow boxes indicate fast tunable MZIs. $N-1$ loops and MZIs are required to implement a universal $N$-mode interferometer.} 
	\label{fig:chain-loop}
\end{figure}
As a compromise between the large potential complexity of spatially-encoded architectures and large potential loss of the dual-loop architecture as $N$ increases, here we consider a chain-loop architecture, making use of temporal encoding and schematically shown in Fig.~\ref{fig:chain-loop}. A chain-loop architecture implementing an $N$-mode universal interferometer consists of $N-1$ MZIs and $N-1$ delay loops. For input time-bins separated by time $\tau$, each loop must implement a delay $\tau$ equal to that of the inner loop of the dual-loop architecture. Note that there is no constraint on the delay $d$ between two adjacent tunable beamsplitters and it can be chosen such that $d \ll \tau$. The number of optical components of the chain-loop architecture increases linearly with the size of the interferometer, whereas the number of optical components of the spatially-encoded architectures increases quadratically. Therefore the control complexity of the chain-loop architecture is reduced with respect to the spatially-encoded architectures. Additionally, there is no need for fast switches in and out of long delay lines in the chain-loop architecture, and therefore the amount of coupling loss is reduced with respect to the dual-loop architecture.

\begin{figure*}[t]
	\includegraphics[scale=0.5]{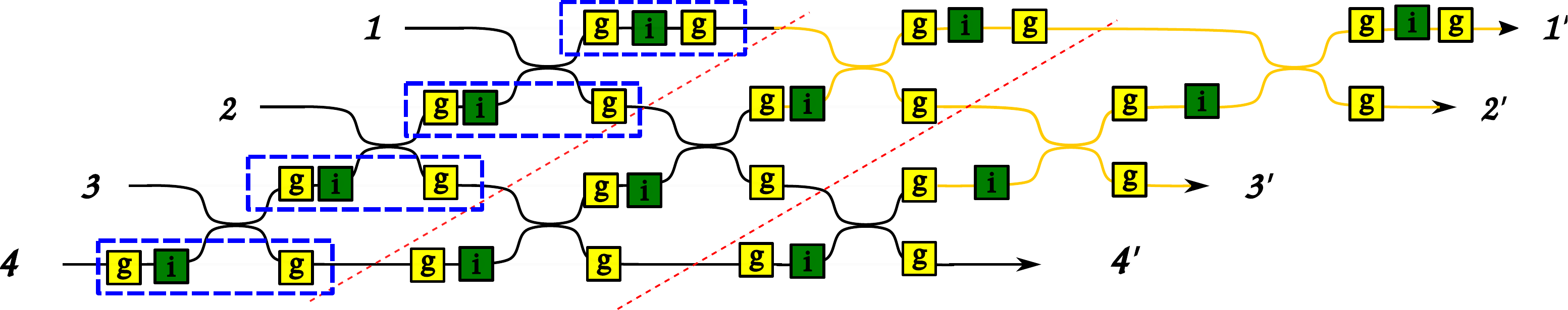}
	\caption{ Implementation of a $4$-mode interferometer with chain-loop architecture. Each time all 4 time-bins travel through a loop, one layer of the Reck scheme is applied. Different layers are separated by red dashed lines. Losses resulting from different optical components are indicated by different boxes: yellow boxes labeled by a ``g'' for MZIs, and green boxes labeled by ``i'' for loops. A detailed procedure outlining how to construct this figure and others like it is given in the main text.}
	\label{fig:chain-loop-loss}
\end{figure*}

Here, each MZI plus loop in the chain implements a layer of beamsplitters of the Reck scheme of Fig.~\ref{fig:spatial-encoding}~(b). With reference to Fig.~\ref{fig:chain-loop}, it is seen that the only reductions in transmission are from the MZIs, which we denote by $\eta_{g}$, and propagation through the delay loops, which we denote by $\eta_{i}$. To detail and help make clear how these transmission factors accumulate, similar to above, we develop a procedure for inserting losses associated with the chain-loop architecture directly on top of a sketch of a Reck scheme (see Fig.~\ref{fig:chain-loop-loss} for the 4-mode case). We use yellow boxes labeled with a ``g'' to indicate MZIs and green boxes labeled with an ``i'' to indicate loops.  

The procedure to mock up a Reck scheme to include chain-loop architecture losses is as follows:
\begin{enumerate}
	\item Sketch the corresponding Reck decomposition for $U(N)$. 
	\item Add dashed diagonal lines (shown in red in Fig.~\ref{fig:chain-loop-loss}) to separate the $N-1$ layers. 
 	\item Add $j-1$ MZIs (shown in yellow in Fig.~\ref{fig:chain-loop-loss}) to each layer $j$ so that each layer contains $N-1$ MZIs. 
	\item Add a gate loss to each of the the two output modes of each MZI ($2\left(N-1\right)^{2}$ in total).
	\item Within each layer, add loop loss to the lower input mode and/or upper output mode of each MZI ($N^{2}$ in total).
	\item Finally, within each layer, add just two gate losses: one on the lower input mode to the lowest MZI, and one on the upper output mode or the uppermost MZI. These two extra losses are needed to push the first time-bin into a loop, and to push the last time-bin out of a loop.
\end{enumerate}

As can be seen in  Fig.~\ref{fig:chain-loop-loss}, the effective transmission here, as for the dual-loop architecture, is the same for each mode and independent of the path taken from input to output. In particular, for an $N=4$ interferometer, each time-bin passes through 6 MZIs and 3 loops. In general, this can be written as
\begin{equation}\label{eq:chain-loop-eff-gate-loss}
\eta_{\text{CL}} = \left(\eta_{g}^2\eta_{i}\right)^{N-1}.
\end{equation}  
Comparing with Eq.~\eqref{eq:double-loop-overall-loss}, we find that the effective transmission of the chain-loop is larger than that of the dual-loop by $\eta_{s}^{2}\eta_{o}$ per layer---exactly the excess loss introduced by the outer loop and switches. However, it is not clear that physical implementations will have the same $\eta_{g}$ or $\eta_{i}$ for each architecture, and thus not immediately obvious, even without introducing questions of scalability and control complexity, which architecture is best for a particular application. Before exploring these questions in more detail in Sec.~\ref{sec:implementations}, we first place our definitions of $\eta$ on more solid mathematical footing.

\section{Quantifying overall transmission }\label{sec:simulation}

Although we can fully describe the lossy interferometer by its transformation on mode operators, it is not immediately obvious how to quantify its effective loss. In the previous section, we proposed using the overall loss connecting input modes $\{i\}$ to output modes $\{i^{\prime}\}$, as a measure of loss in each loop-based architecture. Although this quantity is by no means rigorous, it carries some physical intuition and is easy to calculate. In the present section, we justify this measure by showing that it can be understood as a kind of average loss over all possible optical paths. Before that, we first briefly review how to mathematically model a multi-mode lossy channel.

An ideal universal multiport interferometer is characterized by the unitary transformation
\begin{align}
\hat{\bm{a}}\rightarrow \bm{U}\hat{\bm{a}},
\end{align}
where $\hat{\bm{a}}:=(\hat{a}_1,\ldots,\hat{a}_N)^T$ are the annihilation operators for each mode. Any such unitary can be decomposed into a series of nearest-neighbor two-mode $SU(2)$ gates~\cite{reck1994experimental,clements2016optimal} (physically these gates are usually implemented as MZIs). Let $\bm{R}_{n_i,m_i}$ be the $i^\text{th}$ two-mode unitary acting on modes $n$ and $m$. Then, for a given decomposition, we have
\begin{align}
\bm{U}=\prod_{i}\bm{R}_{n_i,n_i+1}.
\end{align}
For spatial encoding there will be $N(N-1)/2$ such two-mode gates \cite{reck1994experimental,clements2016optimal}. For loop-based architectures (see Fig.~\ref{fig:double-loop-loss} and Fig.~\ref{fig:chain-loop-loss}) there will be $\left(N-1\right)^{2}$ gates because light does not exit immediately after traversing its final required gate.

When loss is included, instead of a unitary transformation, the interferometer is characterized by a multi-mode lossy channel~\cite{giovannetti2015solution}
\begin{align}
\hat{\bm{a}}\rightarrow \bm{A}\hat{\bm{a}} + \sqrt{\bm{I}-\bm{A}\bm{A}^*}\hat{\bm{e}}~,
\end{align}
where $0<\bm{A}\bm{A}^*\leq \bm{I}$, $\bm{A}$ is the process matrix and $\hat{\bm{e}}$ are modes of the environment. The process matrix is constructed by inserting loss matrices in between two-mode unitaries
\begin{align}\label{eq:process-matrix}
\bm{A}= \prod_{i}\bm{L}_{n_i,n_i+1}\bm{R}_{n_i,n_i+1}.
\end{align}
Here $\bm{L}_{n,n+1}$ only acts on the $n^\text{th}$ and $(n+1)^\text{th}$ modes and can be written as
\begin{align}
\bm{L}_{n,n+1} = \begin{pmatrix}
\sqrt{\eta_n} & 0\\
0 & \sqrt{\eta_{n+1}}
\end{pmatrix},
\end{align}
where $\eta_n$ and $\eta_{n+1}$ are the losses suffered by the output modes of the $n$th gate. The values of $\eta_n$ can be found by looking at diagrams like those shown in Fig.~\ref{fig:double-loop-loss} and Fig.~\ref{fig:chain-loop-loss}.

\begin{figure}[t]
	\includegraphics[scale=0.36]{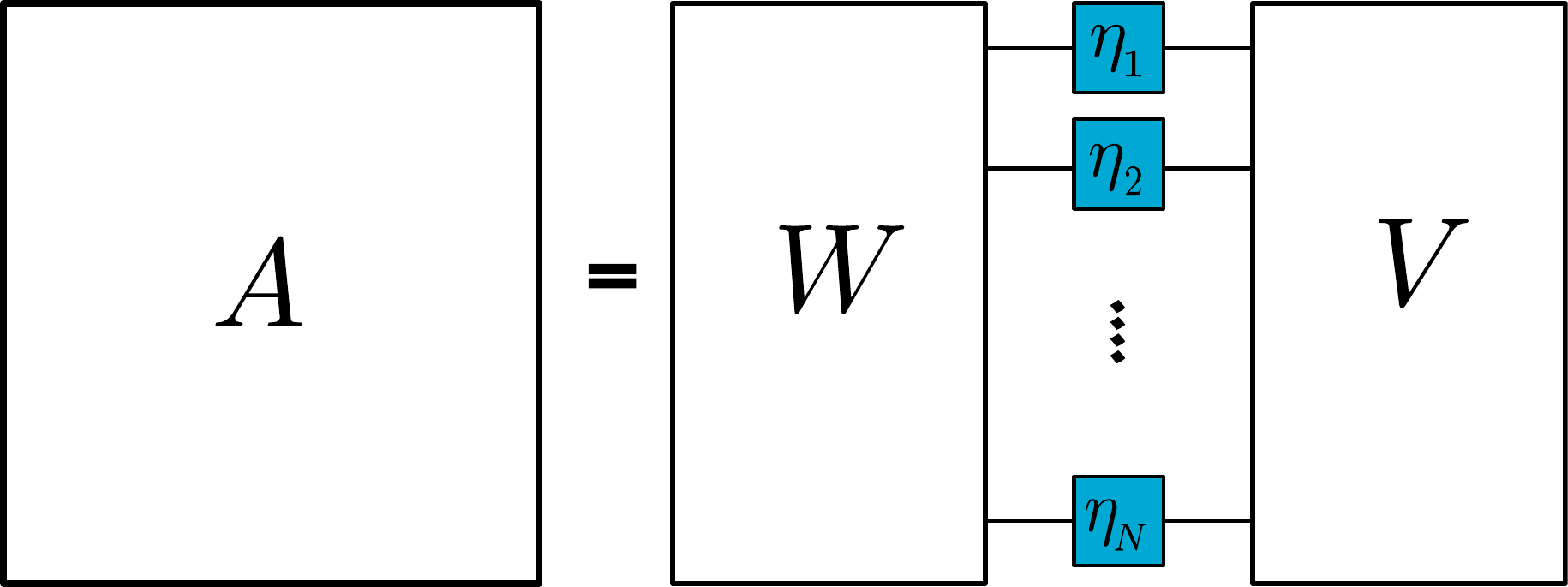}
	\caption{Singular decomposition of the process matrix. Here $\bm{W},\bm{V}$ are unitary matrices and $\brac{\eta_i}$ are $N$ independent lossy channels. Thanks to this decomposition, we can separate the losses from the unitary transformation which is convenient for our loss analysis. \label{fig:svd}}
\end{figure}
From Eq.~\eqref{eq:process-matrix} we can see that losses and unitary transforms are mixed together, making the loss analysis nontrivial. Therefore, to quantify the loss, we follow Ref.~\cite{garcia2017simulating} and first exploit the fact that for any $\bm{A}$, there exists a singular value decomposition $\bm{A}=\bm{V}\bm{\lambda}\bm{W}$ where $\bm{V}$ and $\bm{W}$ are untiary matrices and $\bm{\lambda}=\text{diag}\left(\sqrt{\eta_1},\ldots,\sqrt{\eta_N}\right)$. As we can see from Fig.~\ref{fig:svd}, after the decomposition, losses in different paths are captured by $N$ effective and independent lossy channels and are completely separated from the unitary transformation. Therefore, the average value given by
\begin{align}\label{eq:eta-bar}
\bar{\eta}=\frac{1}{N}\sum_{i=1}^N\eta_i,
\end{align}
could be a good measure of the average loss among all possible paths.
\begin{figure}[h]

	\subfloat[Dual-loop architecture with $\eta_g=0.6, \eta_s=0.75, \eta_i=0.9, \eta_o=0.8^N$.]{%
		\includegraphics[clip,width=\columnwidth]{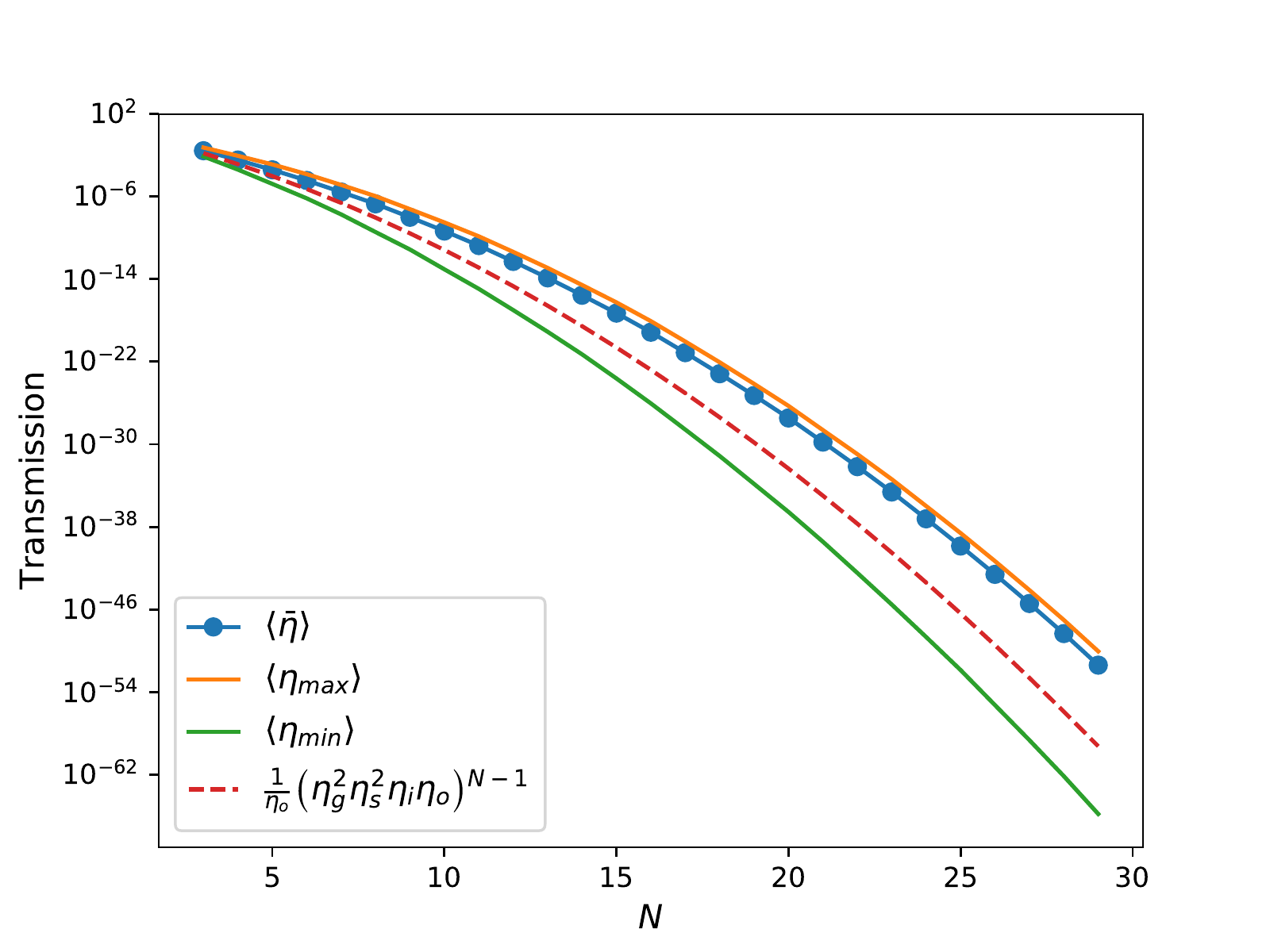}%
	}
	
	\subfloat[Chain-loop architecture with $\eta_g=0.7, \eta_i=0.8$ ]{%
		\includegraphics[clip,width=\columnwidth]{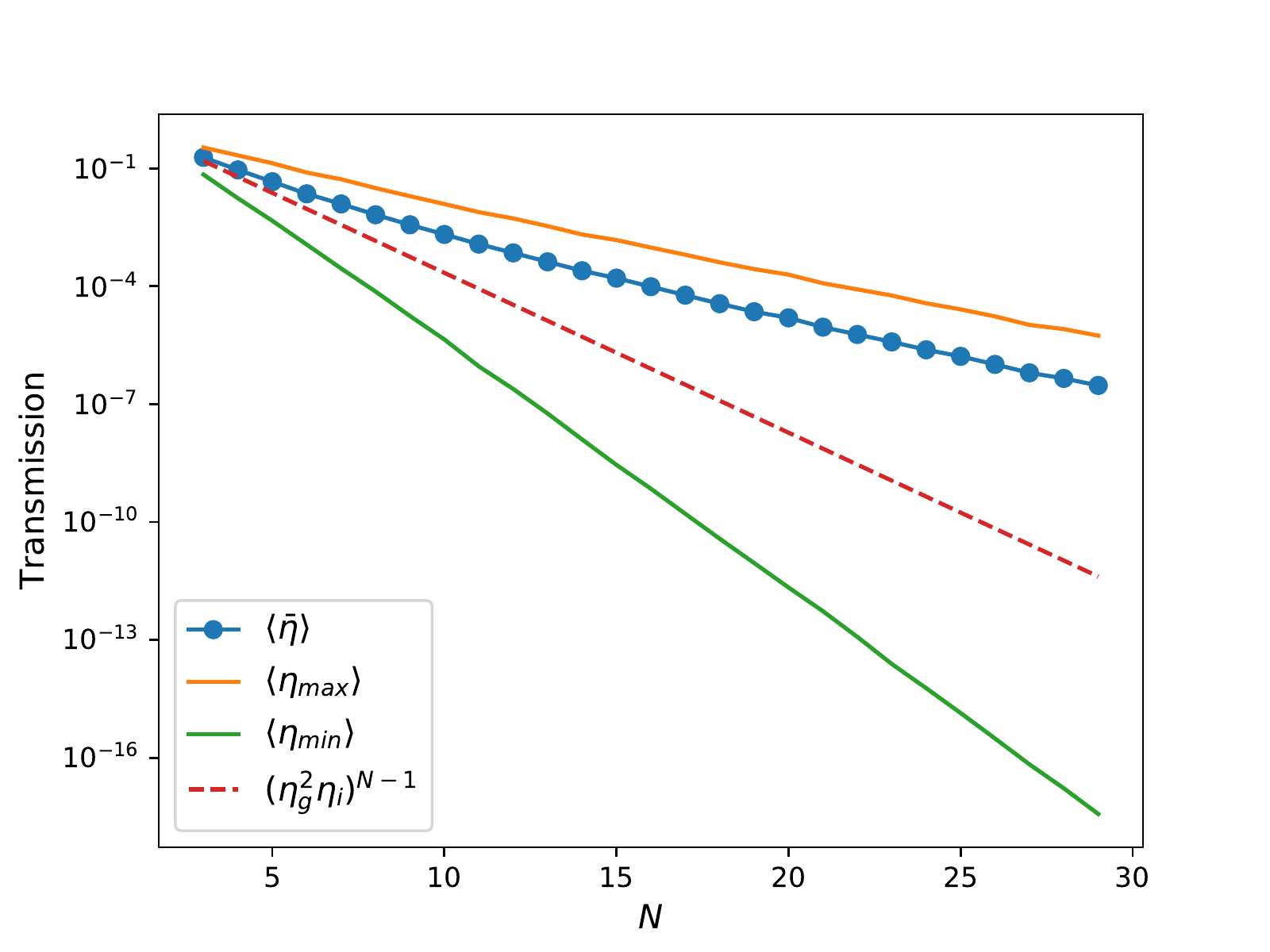}%
	}
	\caption{Simulation of two loop-based architectures. For a fixed number of modes $N$, we randomly draw a unitary transformation $\bm{U}$ from the Haar measure, obtain the corresponding process matrix $\bm{A}$ for each architecture, and determine $\eta_{\max}$, $\eta_{\min}$, $\bar{\eta}$, and $\delta\eta$ from $\bm{A}$'s singular values. We repeat this process 50 times and calculate the mean value of each quantity. In (a) we plot $\eta_{DL}$ of Eq.~\eqref{eq:double-loop-overall-loss} and note that it lies between $\ave{\eta_{\min}}$ and $\ave{\eta_{\max}}$. Similar behavior is found for $\eta_{CL}$ of Eq.~\eqref{eq:chain-loop-eff-gate-loss} in (b).  \label{fig:sim-average}}	
\end{figure}

To better understand and compare the definition of average transmission in Eq.~\eqref{eq:eta-bar} and those in Eq.~\eqref{eq:double-loop-overall-loss} and Eq.~\eqref{eq:chain-loop-eff-gate-loss}, we resort to numerical simulation. For each architecture, we randomly draw unitaries from the Haar measure, calculate the corresponding process matrix, and extract its singular values. We then calculate the maximum, minimum, average, and standard deviation of $\brac{\eta_i}$, denoted by $\eta_{\max}$, $\eta_{\min}$, $\bar{\eta}$, and $\delta\eta$, respectively. In Fig.~\ref{fig:sim-average} we plot the average value of these quantities over 50 random Haar unitaries for several values of $N$. We make two observations from our simulation results: (i) The intuitive definitions of Eqs.~\eqref{eq:double-loop-overall-loss} and~\eqref{eq:chain-loop-eff-gate-loss} always overestimate the loss compared to the average $\ave{\bar{\eta}}$; (ii) $\ave{\bar{\eta}}$ is very close to $\ave{\eta_{\max}}$ and they each scale with $N$ aproximately the same way. Point 2) indicates that losses suffered by most paths across the interferometer are very close to $\ave{\eta_{\max}}$, which suggests certain degree of uniformity. Note that these observations hold throughout extensive numerical simulations and we just present one set of representative parameters in Fig.~\ref{fig:sim-average}. Therefore, $\ave{\bar{\eta}}$ may be a more accurate measure of the average loss than our intuitive definitions. However, as we can see from Fig.~\ref{fig:sim-average}, the transmissions defined in Eq.~\eqref{eq:double-loop-overall-loss} and Eq.~\eqref{eq:chain-loop-eff-gate-loss} always lie between the maximal and minimal values, and are far more close to the center compared to $\ave{\bar{\eta}}$. It is in this sense that we consider the heuristic transmissions defined in Eq.~\eqref{eq:double-loop-overall-loss} and Eq.~\eqref{eq:chain-loop-eff-gate-loss} as valid measures of average loss. In fact, when the loss is not close to uniform (for instance if we allow time-bins to exit the architecture after they complete the last interaction they need to participate in, causing diagrams in Fig.~\ref{fig:double-loop-loss} and Fig.~\ref{fig:chain-loop-loss} to become triangular in shape) our intuitive definitions are perhaps better measures of the average loss than $\ave{\bar{\eta}}$. 

\section{Possible physical implementations}\label{sec:implementations}
In this Section, we consider potential physical implementations of different architectures. Again we first discuss a spatially-encoded architecture, which we later use as a benchmark for temporally-encoded loop-based architectures. We then consider free-space implementations of both loop-based architectures. Although this implementation is not so realistic for the chain-loop when $N$ is large, it is a useful way to give a fair comparison between the two architecture encodings. To overcome problems of size and stability for the chain-loop, we then consider what would be possible with integrated platforms. We expect the chain-loop architecture to benefit most from an integrated design, as it does not require a long delay loop like the dual-loop architecture does. Thus we restrict our consideration of integrated designs to those based on the chain loop architecture. We examine both currently available technology as well as what might be hoped for in the near future. 

Although we seek to compare various multiport interferometer architectures, and are not overly concerned with input state generation or output state detection in this work, it is true that generation and detection components could influence $\tau$. On the state generation side, it is not unreasonable to imagine creating input states separated in time by the repetition period of the laser driving their generation. Taking the repetition rate of a state of the art telecommunications-band mode-locked laser to be in the GHz range, we can safely assume that input state time-bins can be as near to each other as 1~ns. As for detection, while current number-resolving detection systems do not operate at these speeds~\cite{lee2018multi}, we point out that other detection schemes are possible including multiplexing of number-resolving detectors, and technology is always improving. As such, we compare dual-loop and chain-loop architectures both with a minimum $\tau$ of 1 nanosecond.

\subsection{Spatial encoding with integrated platforms}
With its large index of refraction enabling tight bends and small component footprints, silicon is an attractive platform for integrated spatially-encoded interferometers. Recent work implemented 7 layers of MZIs in approximately 1.6~mm~\cite{harris2016large}, giving $L_{\text{MZI}}{\sim}235~\mu\text{m}$. Using a propagation loss of $2.4\text{ dB/cm}$~\cite{harris2016large} and Eq.~\eqref{eq:transmission-spatial-encoding}, we estimate the transmission per layer of this implementation of a spatially-encoded multiport to be $\eta_{g}{\sim}0.987$. If we are more optimistic that one day the propagation loss in silicon-based platforms can be dropped to $0.03\text{ dB/cm}$~\cite{cardenas2009low}, this is further improved to a transmission of $0.998$ per layer. In general, we expect spatially-encoded interferometers on integrated platforms to suffer less loss than those of loop-based architectures, due to the absence of any additional lossy components such as delay loops or switches. However, we note that it becomes increasingly difficult to control the MZIs, not to mention sources and detectors, as the number of modes becomes large. It is at this point that loop-based architectures start to become a better option.

\subsection{Dual-loop architecture in free space}
In free space, the $\tau$ for the dual-loop architecture could be limited by the fast switches necessary for the reconfigurable beamsplitters and/or switches in and out of the outer delay loop. These switches would be implemented in a fast electro- or acousto-optic material. Commercial bulk, free-space coupled, modulators operate at hundreds of MHz~\cite{Qubig}, and so here we consider $\tau=10\text{ ns}$.

The inner loop would likely be implemented in free space, leading to $\eta_{i}{\sim}1$, but the outer loop, if required to be long enough, would likely be implemented in fiber. Thus, in addition to any reduced transmission from the switch in and out of the outer delay loop itself, we must also consider coupling loss in and out of the fiber. The refractive index of a silica fiber at 1560~nm is $n{\sim}1.4$, and so we estimate an outer loop size $N\tau c/n{\sim}N\times 2.1\text{ m}$. We take the propagation loss in the fiber to be 0.2~dB/km, and thus the transmission to be approximately $\eta_{o}{\sim}0.9999^{N}$ or ${\sim}0.9951$ for $N=50$.

Lastly, the dual-loop MZI makes use of two free-space coupled modulators, and each switch makes use of one free-space coupled modulator. These have 98\% transmission~\cite{Qubig}. 
Additionally, the dual-loop architecture will suffer loss when coupling in and out of the outer delay loop fiber of ${\sim}0.3\text{ dB}$. Thus, the MZI has a transmission of $\eta_{g}{\sim}0.9604$, and each switch transmission plus coupling efficiency works out to approximately $\eta_{s}{\sim}0.9146$. These, and all other transmission factors discussed in this Section are summarized in Table~\ref{tab:losses}.% at the start of Subsection \ref{subsec:simulation}.

\subsection{Chain-loop architecture}
\subsubsection{Free space}
As with the free-space implementation of the dual-loop architecture, the chain-loop architecture $\tau$ would be limited by free-space coupled modulators to 10 nanoseconds. In fact, in theory nothing would change and we would have $\eta_{i}{\sim}1$, and $\eta_{g}{\sim}0.9604$. However, because the light would not be refocused by the outer delay loop at regular intervals, it requires additional optical components such as lenses, which would introduce loss. Additionally, phase stability over the entire beam path would ultimately limit the number of modes $N$ that a free-space chain-loop architecture could act on. Thus, while it may have an advantage over a free space implementation of the dual-loop architecture for small $N$, a free space implementation of the chain-loop architecture would become impractical for large $N$. For large $N$, it may benefit from being implemented on an integrated platform.

\subsubsection{Current integrated platform}
To get around issues of phase stability and control complexity, we can imagine placing the entire chain-loop structure on a photonic chip. Propagation losses in an integrated platform are larger than in free space or even fiber, but sizes are smaller, and modulators are faster. In particular low-loss integrated lithium niobate electro-optic modulators operating at several GHz have recently been demonstrated~\cite{wang2018integrated}, and so we can consider lithium niobate as the platform for an integrated chain-loop architecture. 

For $\tau=1\text{ ns}$ and considering the index of refraction in lithium niobate to be $n{\sim}2.2$ at 1560~nm, we estimate a chain-loop architecture delay loop size of $\tau c/n{\sim}14\text{ cm}$.  Taking the propagation loss in the integrated lithium niobate platform to be 2.7~dB/m~\cite{zhang2017monolithic}, we have $\eta_i{\sim}0.9188$. To implement each tunable gate, the chain-loop architecture must use two integrated lithium niobate modulators for each MZI, each with a loss of ${\sim}0.5\text{ dB}$~\cite{wang2018integrated} which yields $\eta_g{\sim}0.7943$.

\begin{figure}[ht!]
\includegraphics[width=4cm]{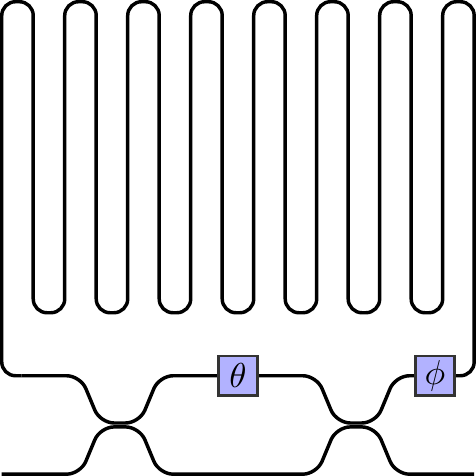}
\caption{ A schematic of an MZI with a delay loop. The MZI consists of two (static) 50:50 beamsplitters and two phase shifters with phases $\theta$ and $\phi$. The phase $\theta$ is used to control the effective transmission and $\phi$ is simply a phase shift. In practice the delay loop would be arranged in a spiral such that it occupies a small area. } 
\label{fig:chain-loop-MZI}
\end{figure}

\begin{figure}[ht!]
\includegraphics[width=8.5cm]{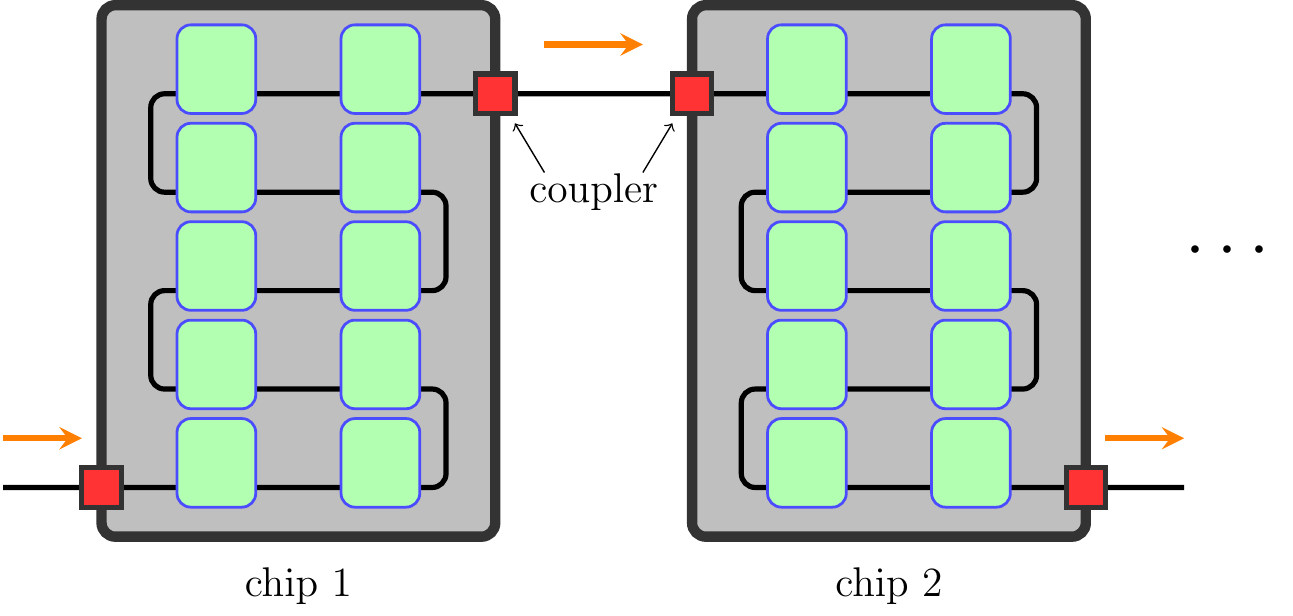}
\caption{ Chain-loop architecture on chip. Two chips (big grey boxes) are schematically shown. Each small green box on the chip is a MZI plus a delay loop, which is shown in Fig.~\ref{fig:chain-loop-MZI}. The small red boxes are couplers that connect different chips. A very large interferometer can be implemented by concatenating more chips.  } 
\label{fig:chain-loop-chip}
\end{figure}

Note that the 14~cm delay loops can be arranged in a spiral configuration to save space (see Fig.~\ref{fig:chain-loop-MZI}), occupying an area on the order of $1\text{ mm}^{2}$. Even so, we note that only so many of them can be made to fit on a single chip. As a way around this, one can imagine connecting multiple chips (see Fig.~\ref{fig:chain-loop-chip} for a schematic) to extend the length of the chain indefinitely. Compared to coupling issues for spatial-encoding schemes when they run out of space on chip, with several modes needing to be coupled from chip to chip, here there is just a single spatial mode to couple. However, we do note that integrated chain-loop architectures will run out of space more quickly, and that any chip-to-chip coupling will increase loss. Estimating this coupling loss requires the knowledge of specifics of experimental implementation, we neglect this coupling loss in the present work.

\subsubsection{Future integrated platform}
In the future we can imagine the integrated lithium niobate modulators to be propagation loss limited. As they are only about 3~mm long~\cite{Mian}, this would result in $\eta_g{\sim}0.9981$. Furthermore, if we imagine propagation losses in integrated lithium niobate to come down to that of certain polished lithium niobate structures~\cite{Mian} or 0.3~dB/m, we would instead have $\eta_i{\sim}0.9906$ and $\eta_g{\sim}0.9998$.

\subsection{Simulation and performance comparison}\label{subsec:simulation}
Finally, we use the definitions of average loss from Sec.~\ref{sec:architectures} and Sec.~\ref{sec:simulation} to compare the performance of loop-based architectures with different physical implementations. We summarize the transmission values discussed in Sec.~\ref{sec:implementations} in Table~\ref{tab:losses}.

\begin{table}[h]
	\begin{tabular}{c|c|c|c|c}
		& DL FS & CL FS & CL I, current & CL I, future\tabularnewline
		\hline 
		Inner loop & ${\sim}1$ & ${\sim}1$ & 0.9188 & 0.9906 \tabularnewline
		\hline 
		Outer loop & $0.9999^{N}$ & N/A & N/A & N/A\tabularnewline
		\hline 
		MZI & 0.9604 & 0.9604 & 0.7943 & 0.9998\tabularnewline
		\hline 
		Switch & 0.91 & N/A & N/A & N/A\tabularnewline
	\end{tabular}
	\caption{Transmission efficiencies through components of each considered loop architecture. DL: dual-loop, CL: chain-loop, FS: free-space, I: integrated. We take values consistent with the literature, though note that the chain-loop architecture in free space is limited in size, and that the future values for the chain-loop architecture in an integrated platform may be overly optimistic.} 
	\label{tab:losses}
\end{table} 

\begin{figure}[h]
	\subfloat[]{%
		\includegraphics[clip,width=\columnwidth]{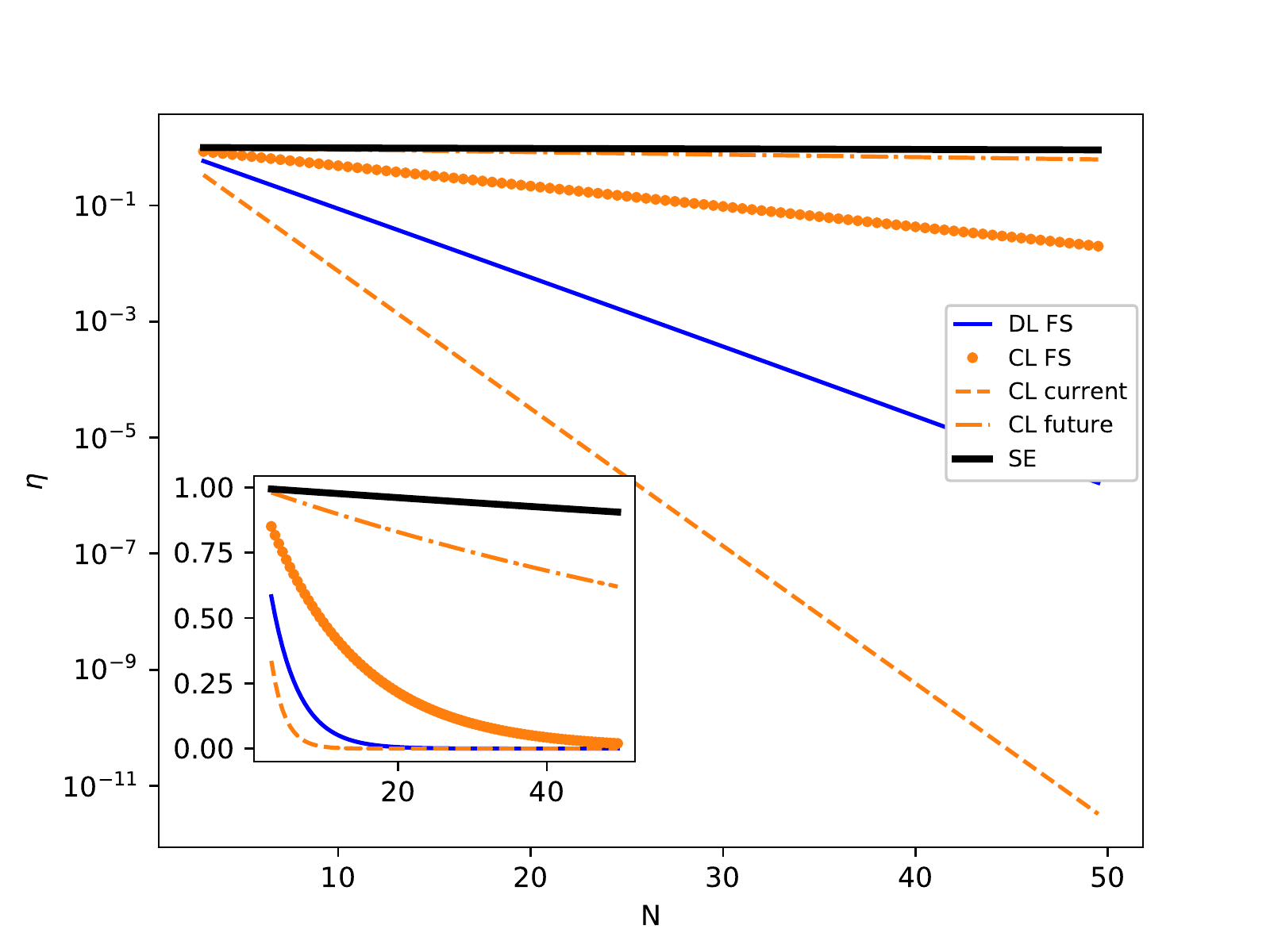}%
	}
	
	\subfloat[]{%
		\includegraphics[clip,width=0.95\columnwidth]{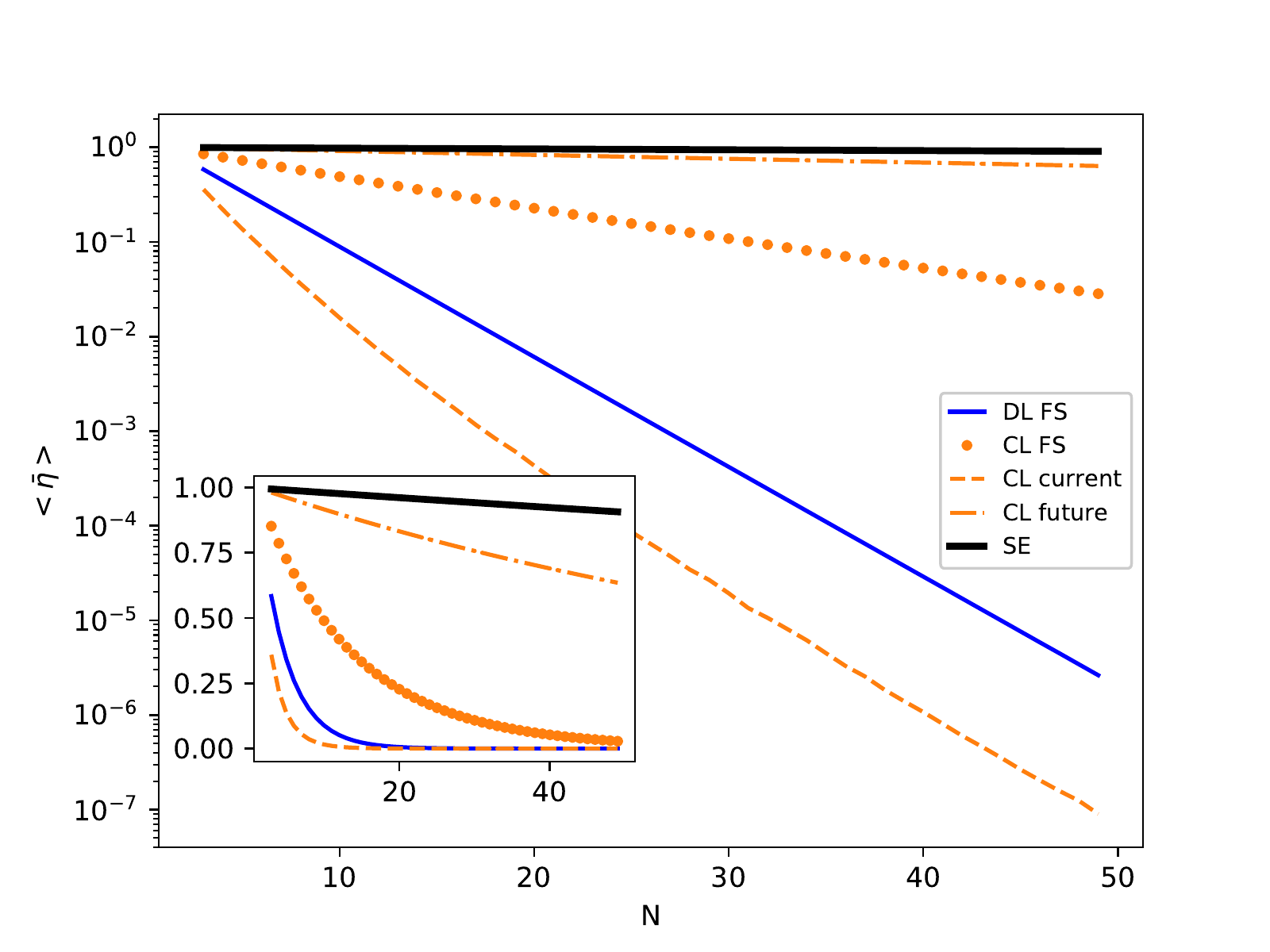}%
	}
	\caption{Performance between loop-based architectures compared using two different definitions of average loss. (a) Logarithmic plot of overall transmission using $\eta_{\text{DL}} = \left(\eta_{g}^2\eta_{s}^2\eta_{i}\eta_{o}\right)^{N-1}\eta_{o}^{-1}$ and $\eta_{\text{CL}}=(\eta_g^2\eta_i)^{N-1}$. (b) Logarithmic plot of average transmission $\ave{\bar{\eta}}$ over 50 random Haar unitaries at each point. In each subplot we plot curves for the dual-loop architecture implemented in free space (DL FS), chain-loop architecture implemented in free space (CL FS), chain-loop architecture implemented on a current integrated platform (CL current), and chain-loop architecture implemented on an optimistic near-future integrated platform (CL future). We also plot $\eta_{SE} = \eta_{g}^{N}$ for a spatially-encoded architecture on an integrated platform (SE) using the numbers provided in Sec.~\ref{sec:simulation}.A. In the insets we plot the same curves on a linear scale.\label{fig:comparision}}
\end{figure}

In Fig.~\ref{fig:comparision} (a) we plot the average losses defined in Eq.~\eqref{eq:double-loop-overall-loss} and Eq.~\eqref{eq:chain-loop-eff-gate-loss}. In (b) we plot $\ave{\bar{\eta}}$ defined in Eq.~\eqref{eq:eta-bar} averaged over 50 random Haar unitaries. We use the data in Tab.~\ref{tab:losses} for both subplots. We find that the difference between these two definitions is only noticeable for sufficiently large loss. For the regime where $\eta>0.1$, these two measures are almost indistinguishable. This is confirmed by looking at the two insets, where the average loss is plotted without rescaling.
Summarizing these figures, it is obvious that for small, controllable, phase-stable interferometers, a chain-loop architecture implemented in free space is likely the best option. However, as the interferometer gets larger, current integrated modulator technology will limit the utility of the chain-loop architecture on chip, leading the dual-loop architecture in free space to likely be the best option. On the other hand, integrated modulator technology does not need to become all that much better for integrated chain-loop architectures to become the lowest loss option. Comparing the loss per layer of each architecture, namely $\eta_{g,\text{DL}}^{2}\eta_{s}^{2}\eta_{i,\text{DL}}\eta_{o}$ and $\eta_{g,\text{CL}}^{2}\eta_{i,\text{CL}}$, we see that, taking $\eta_{i,\text{DL}}{\sim}1$, when 
\begin{equation}
\eta_{g,\text{CL}}>\eta_{g,\text{DL}}\eta_{s}\sqrt{\eta_{o}/\eta_{i,\text{CL}}},
\end{equation}
the chain-loop architecture implemented on a chip will be a competitive option.

\section{Applications}\label{sec:application}
Boson sampling and its variations are probably the most obvious and straightforward applications of loop-based architectures, but are definitely not the only ones. Due to its simplicity, boson sampling is one of the most promising routes to demonstrate quantum supremacy in the near term~\cite{preskill2012quantum,harrow2017quantum}. Indeed, boson sampling motivated the proposal of the dual-loop architecture~\cite{motes2014scalable,motes2015implementing} and many recent demonstrations of integrated linear optics~\cite{carolan2015universal,harris2016large,harris2017quantum,wang2018integrated}.  With proper modification, loop-based architectures can be adapted to universal quantum computation \cite{rohde2015simple,takeda2017universal}. Finally, the utility of loop-based architectures is not limited to quantum information processing, for generalized spatial mode converters~\cite{miller2013self,miller2015perfect} and optical neural networks~\cite{shen2017deep} are examples of classical optical applications. In the following we describe the aforementioned optical applications in more detail. 

\subsection{Boson sampling}
A boson sampler, first introduced by Aaronson and Arkhipov in 2010~\cite{aaronson2011computational}, is a non-universal quantum machine where $N$ single photons are sent into an $M$-mode interferometer, followed by detection with photon-number-resolving (PNR) detectors. A \textsc{BosonSampling} (BS) problem can be formulated based on this setup: given a boson sampler, output a sample from the exact or approximate output photon distribution. It has been proven that this problem cannot be efficiently solved by any classical computer~\cite{aaronson2011computational} unless the polynomial hierarchy collapses, which is believed to be highly unlikely in the computer science community~\cite{sipser2006introduction}. The hardness of BS is rooted in the hardness of calculating permanents, which was proved to be in the \#P-complete complexity class~\cite{valiant1979complexity}. Due to its experimental simplicity compared to fault-tolerant universal quantum computation, boson sampling is one of the most promising candidates to demonstrate quantum supremacy in the near future~\cite{preskill2012quantum,harrow2017quantum}.

The original proposal by Aaaronson and Arkhipov did not consider photon loss, which would be present in any realistic implementation of BS. Therefore, it is paramount to ask if the $\textsc{LossyBosonSampling}$ problem is still hard to simulate. Aaronson and Brod proved that sampling from a lossy boson sampler is still hard if the number of lost photons remains constant as the number of input photons increases~---~an unrealistic situation, considering the nature of optical loss in physical systems.  Later Rahimi-Keshari et al.\ showed that exact lossy BS is always hard under the assumption of perfect detection \cite{rahimi2016sufficient}. Using quite different methods, both Garc\'ia-Patr\'on et al.~\cite{garcia2017simulating} and Oszmaniec and Brod~\cite{oszmaniec2018classical} proved that approximate BS is efficiently simulable if the number of remaining photons is $\O(\sqrt{N})$. These results eliminate most experimental platforms where the photon loss scales exponentially with the system size.

Computational complexity theorists are concerned with the asymptotic behavior of an algorithm, however any experimental demonstration will always be at some finite problem size. Therefore, questions such as how much transmission $\eta$ is enough for a given input photon number $N$ to achieve a quantum advantage is perhaps more meaningful to an experimentalist. In Ref.~\cite{neville2017classical} the authors presented a quantum supremacy region over the $\eta$-$N$ plane. In general, the larger the loss the more photons required to show a quantum advantage. For instance, with 50 input photons, $\eta \gtrsim 0.7$ \cite{neville2017classical} is required to have a non-zero quantum advantage. By looking at Fig.~\ref{fig:comparision}, none of architectures with current experimental parameters can reach this threshold. Even the chain-loop architecture implemented on lithium niobate with optimistic parameters can barely reach this threshold. We note that losses in photon sources and detectors are not included in Fig.~\ref{fig:comparision}. Together with the fact that recently a fast classical algorithm has been proposed~\cite{clifford2018classical}, demonstrations of quantum supremacy using boson sampling are extremely challenging for current technology.

\subsection{Gaussian boson sampling}
Inspired by a variation of BS~\cite{lund2014boson},  Hamilton et al.\ proposed to replace the single photon input states with single-mode squeezed vacuum states. It is this setup we refer to as Gaussian boson sampling (GBS)~\cite{hamilton2017gaussian,kruse2018detailed}. By showing that the output distribution of GBS is directly related to the hafnian of a certain matrix~\cite{barvinok2016combinatorics}, the authors argued that GBS is also a hard problem modulo some conjectures similar to those in the original BS paper \cite{aaronson2011computational}.

Implementing GBS is expected to be less challenging than implementing BS since single-mode squeezed states can be generated deterministically and with high indistinguishability. Another important difference between GBS and BS is that post-selection is not an option anymore. This means that we have to consider the lossy GBS problem where the task is to sample from the output distribution of a Gaussian boson sampler. Although a detailed study of lossy GBS has yet to be presented, it is clear that a low-loss scalable implementation will be required, as the inclusion of loss is expected to make classical simulations of GBS easier. On the other hand, classical algorithms for calculating the hafnian and simulating GBS have recently been tested on the Titan supercomputer~\cite{bjorklund2018faster,gupt2018classical}.  

We note that it appears likely that GBS could have applications beyond simply demonstrating quantum supremacy. Due to the closeness between the hafnian and graph theory, much progress has been made on using GBS to solve graph-related problems including the search of dense subgraphs~\cite{arrazola2018using,arrazola2018quantum}, calculating perfect matchings of graphs~\cite{bradler2018gaussian}, and identifying isomorphic graphs~\cite{bradler2018graph}. Interestingly, a slight extension of GBS (to include displacement at the inputs) can be used to sample from the vibronic spectrum of molecules~\cite{huh2015boson,huh2017vibronic}. Finally, a state-preparation scheme based on GBS was recently proposed and significant improvement on previous methods has been shown \cite{sabapathy2018near} .
\subsection{Universal quantum computation}
\begin{figure}[h]
	\includegraphics[scale=0.6]{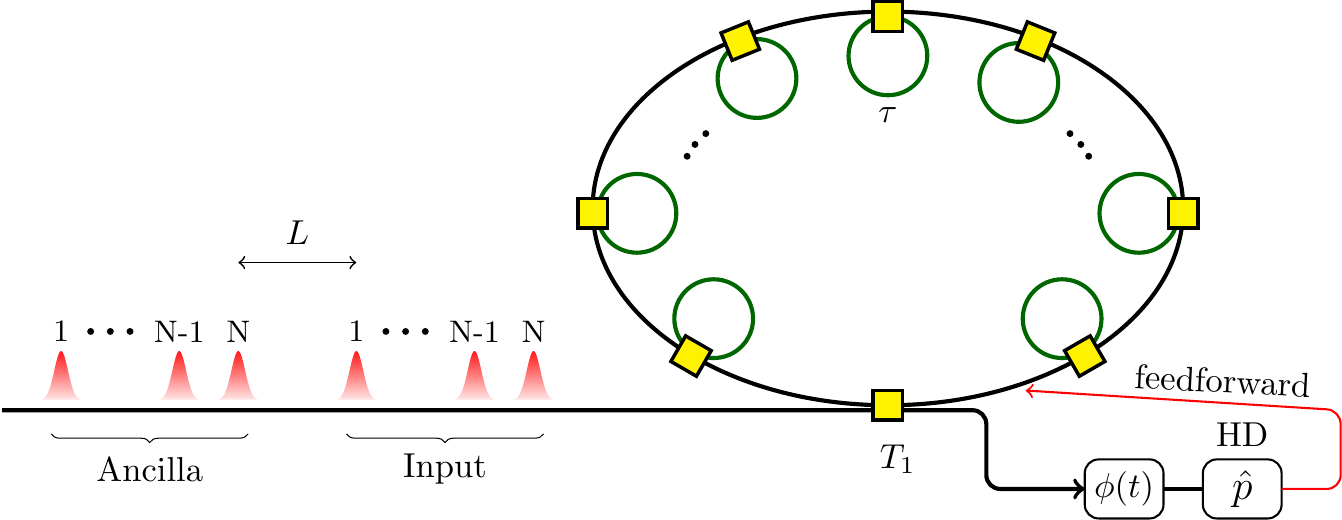}
	\caption{Schematic showing the chain-loop architecture adapted to universal quantum computation. The ancilla modes encode resource states like squeezed states or non-Gaussian states, e.g., cubic phase states.
		The input time-bins are fed into a closed chain to implement a linear unitary transformation, after which they couple with the resource states at the tunable beamsplitter
		$T_1$. One of the outputs of $T_1$ is measured by a homodyne detector and the measurement outcome is used to feed forward another output, thus implementing a
		measurement-induced gate. To make sure the ancilla modes can match with the input modes, the delay between the input and ancilla modes $L$ should be the
		same as the delay of the chain.\label{fig:chain-loop-universal}}
\end{figure}
The dual-loop architecture not only can implement a universal linear interferometer, but also can be adjusted to achieve universal quantum computation  \cite{rohde2015simple,takeda2017universal}.
Similarly, one can straightforwardly adjust the chain-loop architecture to achieve universal quantum computation by introducing 
measurement-induced non-Gaussian gates \cite{GKP2001, Marshall2015, MiyataAdapNonGauss2016, Arzani2017}. The proposed universal quantum computing architecture is shown in Fig. \ref{fig:chain-loop-universal}. A chain which concatenates $N$ delay loops and implements an $N$-mode linear interferometer is closed to form a loop. The purpose is to repeatedly use the chain to implement linear unitary transformations, which are necessary for universal quantum computation. Between these linear transformations, we can implement measurement-induced gates (either squeezing or non-Gaussian gates) by successively feeding resource states (ancilla modes) into the loop, and performing homodyne detection and feedforward.

\subsection{Spatial mode converters}
In addition to the quantum applications listed above, there are several classical applications of a mesh of $SU(2)$ operations. Such meshes, encoded spatially, can be used to implement any linear optical transformation, enabling self-aligning optical mode couplers, combiners, trackers, and separators (see Ref.~\cite{miller2015perfect} and references therein). Similar techniques could readily be applied to the temporally-encoded chain-loop architecture, which should come with reduced control complexity and small loss in the near future.

\subsection{Optical neural networks}
Artificial neural networks are a computing paradigm for machine learning tasks originally inspired by the human brain~\cite{nielsen2018neural}. Rather than relying on a program that explicitly tells it precisely what to do at each step, a neural network ``learns'' from observational data. They have been used successfully for e.g., image and speech recognition. Recently, there has been interest in optical implementations of artificial neural networks~\cite{yeh2004optical,shen2017deep}. These implementations require a multiport photonic interferometer as well as a nonlinear element, such as a saturable absorber, and promise speed and power efficiencies compared to electronic implementations. Already, a few-layer spatially-encoded multiport photonic interferometer implemented in silicon with the nonlinearity simulated on a computer has demonstrated accuracy comparable to a 64-bit computer for a vowel recognition problem~\cite{shen2017deep}. As control complexity increases, the temporally-encoded chain-loop architecture could become a viable candidate for future optical neural networks.

\section{Conclusions}\label{sec:conclusions}
We have presented new loop-based temporally-encoded architecture to implement universal unitary transformation.  We also presented a useful and intuitive framework to evaluate the effective overall loss associated with different physical implementations of temporally-encoded linear multiport interferometers. We justified their use with analytic and numerical calculations demonstrating that, while they may slightly overestimate the ``true" average interferometer loss, they are sufficient to allow for quick and straightforward comparisons, enabling faster progress when comparing interferometric architectures. A procedure is developed for connecting the losses associated with either a dual-loop or chain-loop architecture to the more conventional spatially-encoded architecture, furnishing a clear mapping between spatial and time-encoded decompositions. We expect these results to be useful for targeting the optimal architecture for a diverse range of applications, both quantum and classical. We showed that our chain-loop architecture could outperform dual-loop systems pending realistic near-term improvements to nanophotonic lithium niobate platforms, further motivating experimental efforts to drive down losses in such platforms.

%we have presented useful and intuitive expressions for the effective overall loss associated with different physical implementations of temporally-encoded linear multiport photonic interferometers. We have justified their use on solid mathematical results backed up with numerical simulations demonstrating that, while they may slightly overestimate a ``true'' average loss per interferometer, they are excellent proxies and therefore allow for quick and straightforward comparisons. In conjunction, we also developed a procedure for connecting the losses associated with either a dual-loop or chain-loop architecture to a more conventional spatially-encoded architecture. We expect these results to be useful in helping to target the optimal architecture for a given application. We used the expressions developed to calculate that integrated lithium niobate chain-loop architectures could outperform dual-loop architectures in the near future.

\acknowledgements{The authors thank Blair Morrison, William~R. Clements, Jonathan Lavoie, Nicol\'as Quesada, Matthew Collins, Saleh~Rahimi-Keshari, Casey Myers and Christian Weedbrook for useful discussions. }

\bibliography{loops}

\begin{thebibliography}{50}%
\makeatletter
\providecommand \@ifxundefined [1]{%
 \@ifx{#1\undefined}
}%
\providecommand \@ifnum [1]{%
 \ifnum #1\expandafter \@firstoftwo
 \else \expandafter \@secondoftwo
 \fi
}%
\providecommand \@ifx [1]{%
 \ifx #1\expandafter \@firstoftwo
 \else \expandafter \@secondoftwo
 \fi
}%
\providecommand \natexlab [1]{#1}%
\providecommand \enquote  [1]{``#1''}%
\providecommand \bibnamefont  [1]{#1}%
\providecommand \bibfnamefont [1]{#1}%
\providecommand \citenamefont [1]{#1}%
\providecommand \href@noop [0]{\@secondoftwo}%
\providecommand \href [0]{\begingroup \@sanitize@url \@href}%
\providecommand \@href[1]{\@@startlink{#1}\@@href}%
\providecommand \@@href[1]{\endgroup#1\@@endlink}%
\providecommand \@sanitize@url [0]{\catcode `\\12\catcode `\$12\catcode
  `\&12\catcode `\#12\catcode `\^12\catcode `\_12\catcode `\%12\relax}%
\providecommand \@@startlink[1]{}%
\providecommand \@@endlink[0]{}%
\providecommand \url  [0]{\begingroup\@sanitize@url \@url }%
\providecommand \@url [1]{\endgroup\@href {#1}{\urlprefix }}%
\providecommand \urlprefix  [0]{URL }%
\providecommand \Eprint [0]{\href }%
\providecommand \doibase [0]{http://dx.doi.org/}%
\providecommand \selectlanguage [0]{\@gobble}%
\providecommand \bibinfo  [0]{\@secondoftwo}%
\providecommand \bibfield  [0]{\@secondoftwo}%
\providecommand \translation [1]{[#1]}%
\providecommand \BibitemOpen [0]{}%
\providecommand \bibitemStop [0]{}%
\providecommand \bibitemNoStop [0]{.\EOS\space}%
\providecommand \EOS [0]{\spacefactor3000\relax}%
\providecommand \BibitemShut  [1]{\csname bibitem#1\endcsname}%
\let\auto@bib@innerbib\@empty
%</preamble>
\bibitem [{\citenamefont {Motes}\ \emph {et~al.}(2014)\citenamefont {Motes},
  \citenamefont {Gilchrist}, \citenamefont {Dowling},\ and\ \citenamefont
  {Rohde}}]{motes2014scalable}%
  \BibitemOpen
  \bibfield  {author} {\bibinfo {author} {\bibfnamefont {K.~R.}\ \bibnamefont
  {Motes}}, \bibinfo {author} {\bibfnamefont {A.}~\bibnamefont {Gilchrist}},
  \bibinfo {author} {\bibfnamefont {J.~P.}\ \bibnamefont {Dowling}}, \ and\
  \bibinfo {author} {\bibfnamefont {P.~P.}\ \bibnamefont {Rohde}},\ }\bibfield
  {title} {\enquote {\bibinfo {title} {Scalable boson sampling with time-bin
  encoding using a loop-based architecture},}\ }\href@noop {} {\bibfield
  {journal} {\bibinfo  {journal} {Phys. Rev. Lett.}\ }\textbf {\bibinfo
  {volume} {113}},\ \bibinfo {pages} {120501} (\bibinfo {year}
  {2014})}\BibitemShut {NoStop}%
\bibitem [{\citenamefont {Motes}\ \emph {et~al.}(2015)\citenamefont {Motes},
  \citenamefont {Dowling}, \citenamefont {Gilchrist},\ and\ \citenamefont
  {Rohde}}]{motes2015implementing}%
  \BibitemOpen
  \bibfield  {author} {\bibinfo {author} {\bibfnamefont {K.~R.}\ \bibnamefont
  {Motes}}, \bibinfo {author} {\bibfnamefont {J.~P.}\ \bibnamefont {Dowling}},
  \bibinfo {author} {\bibfnamefont {A.}~\bibnamefont {Gilchrist}}, \ and\
  \bibinfo {author} {\bibfnamefont {P.~P.}\ \bibnamefont {Rohde}},\ }\bibfield
  {title} {\enquote {\bibinfo {title} {Implementing bosonsampling with time-bin
  encoding: Analysis of loss, mode mismatch, and time jitter},}\ }\href@noop {}
  {\bibfield  {journal} {\bibinfo  {journal} {Phys. Rev. A}\ }\textbf {\bibinfo
  {volume} {92}},\ \bibinfo {pages} {052319} (\bibinfo {year}
  {2015})}\BibitemShut {NoStop}%
\bibitem [{\citenamefont {Rohde}(2015)}]{rohde2015simple}%
  \BibitemOpen
  \bibfield  {author} {\bibinfo {author} {\bibfnamefont {P.~P.}\ \bibnamefont
  {Rohde}},\ }\bibfield  {title} {\enquote {\bibinfo {title} {Simple scheme for
  universal linear-optics quantum computing with constant experimental
  complexity using fiber loops},}\ }\href@noop {} {\bibfield  {journal}
  {\bibinfo  {journal} {Phys. Rev. A}\ }\textbf {\bibinfo {volume} {91}},\
  \bibinfo {pages} {012306} (\bibinfo {year} {2015})}\BibitemShut {NoStop}%
\bibitem [{\citenamefont {He}\ \emph {et~al.}(2017)\citenamefont {He},
  \citenamefont {Ding}, \citenamefont {Su}, \citenamefont {Huang},
  \citenamefont {Qin}, \citenamefont {Wang}, \citenamefont {Unsleber},
  \citenamefont {Chen}, \citenamefont {Wang}, \citenamefont {He}, \citenamefont
  {Wang}, \citenamefont {Zhang}, \citenamefont {Chen}, \citenamefont
  {Schneider}, \citenamefont {Kamp}, \citenamefont {You}, \citenamefont {Wang},
  \citenamefont {H{\"o}fling}, \citenamefont {Lu},\ and\ \citenamefont
  {Pan}}]{he2017time}%
  \BibitemOpen
  \bibfield  {author} {\bibinfo {author} {\bibfnamefont {Y.}~\bibnamefont
  {He}}, \bibinfo {author} {\bibfnamefont {X.}~\bibnamefont {Ding}}, \bibinfo
  {author} {\bibfnamefont {Z.-E.}\ \bibnamefont {Su}}, \bibinfo {author}
  {\bibfnamefont {H.-L.}\ \bibnamefont {Huang}}, \bibinfo {author}
  {\bibfnamefont {J.}~\bibnamefont {Qin}}, \bibinfo {author} {\bibfnamefont
  {C.}~\bibnamefont {Wang}}, \bibinfo {author} {\bibfnamefont {S.}~\bibnamefont
  {Unsleber}}, \bibinfo {author} {\bibfnamefont {C.}~\bibnamefont {Chen}},
  \bibinfo {author} {\bibfnamefont {H.}~\bibnamefont {Wang}}, \bibinfo {author}
  {\bibfnamefont {Y.-M.}\ \bibnamefont {He}}, \bibinfo {author} {\bibfnamefont
  {X.-L.}\ \bibnamefont {Wang}}, \bibinfo {author} {\bibfnamefont {W.-J.}\
  \bibnamefont {Zhang}}, \bibinfo {author} {\bibfnamefont {S.-J.}\ \bibnamefont
  {Chen}}, \bibinfo {author} {\bibfnamefont {C.}~\bibnamefont {Schneider}},
  \bibinfo {author} {\bibfnamefont {M.}~\bibnamefont {Kamp}}, \bibinfo {author}
  {\bibfnamefont {L.-X.}\ \bibnamefont {You}}, \bibinfo {author} {\bibfnamefont
  {Z.}~\bibnamefont {Wang}}, \bibinfo {author} {\bibfnamefont {S.}~\bibnamefont
  {H{\"o}fling}}, \bibinfo {author} {\bibfnamefont {C.-Y.}\ \bibnamefont {Lu}},
  \ and\ \bibinfo {author} {\bibfnamefont {J.-W.}\ \bibnamefont {Pan}},\
  }\bibfield  {title} {\enquote {\bibinfo {title} {Time-bin-encoded boson
  sampling with a single-photon device},}\ }\href@noop {} {\bibfield  {journal}
  {\bibinfo  {journal} {Phys. Rev. Lett.}\ }\textbf {\bibinfo {volume} {118}},\
  \bibinfo {pages} {190501} (\bibinfo {year} {2017})}\BibitemShut {NoStop}%
\bibitem [{\citenamefont {Takeda}\ and\ \citenamefont
  {Furusawa}(2017)}]{takeda2017universal}%
  \BibitemOpen
  \bibfield  {author} {\bibinfo {author} {\bibfnamefont {S.}~\bibnamefont
  {Takeda}}\ and\ \bibinfo {author} {\bibfnamefont {A.}~\bibnamefont
  {Furusawa}},\ }\bibfield  {title} {\enquote {\bibinfo {title} {Universal
  quantum computing with measurement-induced continuous-variable gate sequence
  in a loop-based architecture},}\ }\href@noop {} {\bibfield  {journal}
  {\bibinfo  {journal} {Phys. Rev. Lett.}\ }\textbf {\bibinfo {volume} {119}},\
  \bibinfo {pages} {120504} (\bibinfo {year} {2017})}\BibitemShut {NoStop}%
\bibitem [{\citenamefont {Reck}\ \emph {et~al.}(1994)\citenamefont {Reck},
  \citenamefont {Zeilinger}, \citenamefont {Bernstein},\ and\ \citenamefont
  {Bertani}}]{reck1994experimental}%
  \BibitemOpen
  \bibfield  {author} {\bibinfo {author} {\bibfnamefont {M.}~\bibnamefont
  {Reck}}, \bibinfo {author} {\bibfnamefont {A.}~\bibnamefont {Zeilinger}},
  \bibinfo {author} {\bibfnamefont {H.~J.}\ \bibnamefont {Bernstein}}, \ and\
  \bibinfo {author} {\bibfnamefont {P.}~\bibnamefont {Bertani}},\ }\bibfield
  {title} {\enquote {\bibinfo {title} {Experimental realization of any discrete
  unitary operator},}\ }\href@noop {} {\bibfield  {journal} {\bibinfo
  {journal} {Phys. Rev. Lett}\ }\textbf {\bibinfo {volume} {73}},\ \bibinfo
  {pages} {58} (\bibinfo {year} {1994})}\BibitemShut {NoStop}%
\bibitem [{\citenamefont {Clements}\ \emph {et~al.}(2016)\citenamefont
  {Clements}, \citenamefont {Humphreys}, \citenamefont {Metcalf}, \citenamefont
  {Kolthammer},\ and\ \citenamefont {Walmsley}}]{clements2016optimal}%
  \BibitemOpen
  \bibfield  {author} {\bibinfo {author} {\bibfnamefont {W.~R.}\ \bibnamefont
  {Clements}}, \bibinfo {author} {\bibfnamefont {P.~C.}\ \bibnamefont
  {Humphreys}}, \bibinfo {author} {\bibfnamefont {B.~J.}\ \bibnamefont
  {Metcalf}}, \bibinfo {author} {\bibfnamefont {W.~S.}\ \bibnamefont
  {Kolthammer}}, \ and\ \bibinfo {author} {\bibfnamefont {I.~A.}\ \bibnamefont
  {Walmsley}},\ }\bibfield  {title} {\enquote {\bibinfo {title} {Optimal design
  for universal multiport interferometers},}\ }\href@noop {} {\bibfield
  {journal} {\bibinfo  {journal} {Optica}\ }\textbf {\bibinfo {volume} {3}},\
  \bibinfo {pages} {1460--1465} (\bibinfo {year} {2016})}\BibitemShut {NoStop}%
\bibitem [{\citenamefont {de~Guise}\ \emph {et~al.}(2018)\citenamefont
  {de~Guise}, \citenamefont {Di~Matteo},\ and\ \citenamefont
  {S{\'a}nchez-Soto}}]{de2018simple}%
  \BibitemOpen
  \bibfield  {author} {\bibinfo {author} {\bibfnamefont {H.}~\bibnamefont
  {de~Guise}}, \bibinfo {author} {\bibfnamefont {O.}~\bibnamefont {Di~Matteo}},
  \ and\ \bibinfo {author} {\bibfnamefont {L.~L.}\ \bibnamefont
  {S{\'a}nchez-Soto}},\ }\bibfield  {title} {\enquote {\bibinfo {title} {Simple
  factorization of unitary transformations},}\ }\href@noop {} {\bibfield
  {journal} {\bibinfo  {journal} {Phys. Rev. A}\ }\textbf {\bibinfo {volume}
  {97}},\ \bibinfo {pages} {022328} (\bibinfo {year} {2018})}\BibitemShut
  {NoStop}%
\bibitem [{\citenamefont {Harris}\ \emph {et~al.}(2016)\citenamefont {Harris},
  \citenamefont {Bunandar}, \citenamefont {Pant}, \citenamefont {Steinbrecher},
  \citenamefont {Mower}, \citenamefont {Prabhu}, \citenamefont {Baehr-Jones},
  \citenamefont {Hochberg},\ and\ \citenamefont {Englund}}]{harris2016large}%
  \BibitemOpen
  \bibfield  {author} {\bibinfo {author} {\bibfnamefont {N.~C.}\ \bibnamefont
  {Harris}}, \bibinfo {author} {\bibfnamefont {D.}~\bibnamefont {Bunandar}},
  \bibinfo {author} {\bibfnamefont {M.}~\bibnamefont {Pant}}, \bibinfo {author}
  {\bibfnamefont {G.~R.}\ \bibnamefont {Steinbrecher}}, \bibinfo {author}
  {\bibfnamefont {J.}~\bibnamefont {Mower}}, \bibinfo {author} {\bibfnamefont
  {M.}~\bibnamefont {Prabhu}}, \bibinfo {author} {\bibfnamefont
  {T.}~\bibnamefont {Baehr-Jones}}, \bibinfo {author} {\bibfnamefont
  {M.}~\bibnamefont {Hochberg}}, \ and\ \bibinfo {author} {\bibfnamefont
  {D.}~\bibnamefont {Englund}},\ }\bibfield  {title} {\enquote {\bibinfo
  {title} {Large-scale quantum photonic circuits in silicon},}\ }\href@noop {}
  {\bibfield  {journal} {\bibinfo  {journal} {Nanophotonics}\ }\textbf
  {\bibinfo {volume} {5}},\ \bibinfo {pages} {456--468} (\bibinfo {year}
  {2016})}\BibitemShut {NoStop}%
\bibitem [{\citenamefont {Harris}\ \emph {et~al.}(2017)\citenamefont {Harris},
  \citenamefont {Steinbrecher}, \citenamefont {Prabhu}, \citenamefont {Lahini},
  \citenamefont {Mower}, \citenamefont {Bunandar}, \citenamefont {Chen},
  \citenamefont {Wong}, \citenamefont {Baehr-Jones}, \citenamefont {Hochberg},
  \citenamefont {Lloyd},\ and\ \citenamefont {Englund}}]{harris2017quantum}%
  \BibitemOpen
  \bibfield  {author} {\bibinfo {author} {\bibfnamefont {N.~C.}\ \bibnamefont
  {Harris}}, \bibinfo {author} {\bibfnamefont {G.~R.}\ \bibnamefont
  {Steinbrecher}}, \bibinfo {author} {\bibfnamefont {M.}~\bibnamefont
  {Prabhu}}, \bibinfo {author} {\bibfnamefont {Y.}~\bibnamefont {Lahini}},
  \bibinfo {author} {\bibfnamefont {J.}~\bibnamefont {Mower}}, \bibinfo
  {author} {\bibfnamefont {D.}~\bibnamefont {Bunandar}}, \bibinfo {author}
  {\bibfnamefont {C.}~\bibnamefont {Chen}}, \bibinfo {author} {\bibfnamefont
  {F.~N.~C.}\ \bibnamefont {Wong}}, \bibinfo {author} {\bibfnamefont
  {T.}~\bibnamefont {Baehr-Jones}}, \bibinfo {author} {\bibfnamefont
  {M.}~\bibnamefont {Hochberg}}, \bibinfo {author} {\bibfnamefont {Seth}\
  \bibnamefont {Lloyd}}, \ and\ \bibinfo {author} {\bibfnamefont
  {D.}~\bibnamefont {Englund}},\ }\bibfield  {title} {\enquote {\bibinfo
  {title} {Quantum transport simulations in a programmable nanophotonic
  processor},}\ }\href@noop {} {\bibfield  {journal} {\bibinfo  {journal} {Nat.
  Photonics}\ }\textbf {\bibinfo {volume} {11}},\ \bibinfo {pages} {447}
  (\bibinfo {year} {2017})}\BibitemShut {NoStop}%
\bibitem [{\citenamefont {Giovannetti}\ \emph {et~al.}(2015)\citenamefont
  {Giovannetti}, \citenamefont {Holevo},\ and\ \citenamefont
  {Garc\'ia-Patr\'on}}]{giovannetti2015solution}%
  \BibitemOpen
  \bibfield  {author} {\bibinfo {author} {\bibfnamefont {V.}~\bibnamefont
  {Giovannetti}}, \bibinfo {author} {\bibfnamefont {A.~S.}\ \bibnamefont
  {Holevo}}, \ and\ \bibinfo {author} {\bibfnamefont {R.}~\bibnamefont
  {Garc\'ia-Patr\'on}},\ }\bibfield  {title} {\enquote {\bibinfo {title} {A
  solution of {G}aussian optimizer conjecture for quantum channels},}\
  }\href@noop {} {\bibfield  {journal} {\bibinfo  {journal} {Commun. Math.
  Phys.}\ }\textbf {\bibinfo {volume} {334}},\ \bibinfo {pages} {1553--1571}
  (\bibinfo {year} {2015})}\BibitemShut {NoStop}%
\bibitem [{\citenamefont {Garc{\'\i}a-Patr{\'o}n}\ \emph
  {et~al.}(2017)\citenamefont {Garc{\'\i}a-Patr{\'o}n}, \citenamefont
  {Renema},\ and\ \citenamefont {Shchesnovich}}]{garcia2017simulating}%
  \BibitemOpen
  \bibfield  {author} {\bibinfo {author} {\bibfnamefont {R.}~\bibnamefont
  {Garc{\'\i}a-Patr{\'o}n}}, \bibinfo {author} {\bibfnamefont {J.~J.}\
  \bibnamefont {Renema}}, \ and\ \bibinfo {author} {\bibfnamefont
  {V.}~\bibnamefont {Shchesnovich}},\ }\bibfield  {title} {\enquote {\bibinfo
  {title} {Simulating boson sampling in lossy architectures},}\ }\href@noop {}
  {\bibfield  {journal} {\bibinfo  {journal} {arXiv preprint arXiv:1712.10037}\
  } (\bibinfo {year} {2017})}\BibitemShut {NoStop}%
\bibitem [{\citenamefont {Lee}\ \emph {et~al.}(2018)\citenamefont {Lee},
  \citenamefont {Shen}, \citenamefont {Cer{\`e}}, \citenamefont {Gerrits},
  \citenamefont {Lita}, \citenamefont {Nam},\ and\ \citenamefont
  {Kurtsiefer}}]{lee2018multi}%
  \BibitemOpen
  \bibfield  {author} {\bibinfo {author} {\bibfnamefont {J.}~\bibnamefont
  {Lee}}, \bibinfo {author} {\bibfnamefont {L.}~\bibnamefont {Shen}}, \bibinfo
  {author} {\bibfnamefont {A.}~\bibnamefont {Cer{\`e}}}, \bibinfo {author}
  {\bibfnamefont {T.}~\bibnamefont {Gerrits}}, \bibinfo {author} {\bibfnamefont
  {A.~E.}\ \bibnamefont {Lita}}, \bibinfo {author} {\bibfnamefont {S.~W.}\
  \bibnamefont {Nam}}, \ and\ \bibinfo {author} {\bibfnamefont
  {C.}~\bibnamefont {Kurtsiefer}},\ }\bibfield  {title} {\enquote {\bibinfo
  {title} {Multi-pulse fitting of transition edge sensor signals from a
  near-infrared continuous-wave source},}\ }\href@noop {} {\bibfield  {journal}
  {\bibinfo  {journal} {arXiv preprint arXiv:1808.08830}\ } (\bibinfo {year}
  {2018})}\BibitemShut {NoStop}%
\bibitem [{\citenamefont {Cardenas}\ \emph {et~al.}(2009)\citenamefont
  {Cardenas}, \citenamefont {Poitras}, \citenamefont {Robinson}, \citenamefont
  {Preston}, \citenamefont {Chen},\ and\ \citenamefont
  {Lipson}}]{cardenas2009low}%
  \BibitemOpen
  \bibfield  {author} {\bibinfo {author} {\bibfnamefont {J.}~\bibnamefont
  {Cardenas}}, \bibinfo {author} {\bibfnamefont {C.~B.}\ \bibnamefont
  {Poitras}}, \bibinfo {author} {\bibfnamefont {J.~T.}\ \bibnamefont
  {Robinson}}, \bibinfo {author} {\bibfnamefont {K.}~\bibnamefont {Preston}},
  \bibinfo {author} {\bibfnamefont {L.}~\bibnamefont {Chen}}, \ and\ \bibinfo
  {author} {\bibfnamefont {M.}~\bibnamefont {Lipson}},\ }\bibfield  {title}
  {\enquote {\bibinfo {title} {Low loss etchless silicon photonic
  waveguides},}\ }\href@noop {} {\bibfield  {journal} {\bibinfo  {journal}
  {Opt. Express}\ }\textbf {\bibinfo {volume} {17}},\ \bibinfo {pages}
  {4752--4757} (\bibinfo {year} {2009})}\BibitemShut {NoStop}%
\bibitem [{Qub()}]{Qubig}%
  \BibitemOpen
  \href@noop {} {\enquote {\bibinfo {title} {Qubig free space electro-optic
  modulators},}\ }\bibinfo {howpublished}
  {\url{https://www.qubig.com/products/electro-optic-modulators-230/phase-shifters/ps-swir.204.html}},\
  \bibinfo {note} {accessed: 2018-12-06}\BibitemShut {NoStop}%
\bibitem [{\citenamefont {Wang}\ \emph {et~al.}(2018)\citenamefont {Wang},
  \citenamefont {Zhang}, \citenamefont {Chen}, \citenamefont {Bertrand},
  \citenamefont {Shams-Ansari}, \citenamefont {Chandrasekhar}, \citenamefont
  {Winzer},\ and\ \citenamefont {Lon{\v{c}}ar}}]{wang2018integrated}%
  \BibitemOpen
  \bibfield  {author} {\bibinfo {author} {\bibfnamefont {C.}~\bibnamefont
  {Wang}}, \bibinfo {author} {\bibfnamefont {M.}~\bibnamefont {Zhang}},
  \bibinfo {author} {\bibfnamefont {X.}~\bibnamefont {Chen}}, \bibinfo {author}
  {\bibfnamefont {M.}~\bibnamefont {Bertrand}}, \bibinfo {author}
  {\bibfnamefont {A.}~\bibnamefont {Shams-Ansari}}, \bibinfo {author}
  {\bibfnamefont {S.}~\bibnamefont {Chandrasekhar}}, \bibinfo {author}
  {\bibfnamefont {P.}~\bibnamefont {Winzer}}, \ and\ \bibinfo {author}
  {\bibfnamefont {M.}~\bibnamefont {Lon{\v{c}}ar}},\ }\bibfield  {title}
  {\enquote {\bibinfo {title} {Integrated lithium niobate electro-optic
  modulators operating at cmos-compatible voltages},}\ }\href@noop {}
  {\bibfield  {journal} {\bibinfo  {journal} {Nature}\ }\textbf {\bibinfo
  {volume} {562}},\ \bibinfo {pages} {101} (\bibinfo {year}
  {2018})}\BibitemShut {NoStop}%
\bibitem [{\citenamefont {Zhang}\ \emph {et~al.}(2017)\citenamefont {Zhang},
  \citenamefont {Wang}, \citenamefont {Cheng}, \citenamefont {Shams-Ansari},\
  and\ \citenamefont {Lon\v{c}ar}}]{zhang2017monolithic}%
  \BibitemOpen
  \bibfield  {author} {\bibinfo {author} {\bibfnamefont {M.}~\bibnamefont
  {Zhang}}, \bibinfo {author} {\bibfnamefont {C.}~\bibnamefont {Wang}},
  \bibinfo {author} {\bibfnamefont {R.}~\bibnamefont {Cheng}}, \bibinfo
  {author} {\bibfnamefont {A.}~\bibnamefont {Shams-Ansari}}, \ and\ \bibinfo
  {author} {\bibfnamefont {M.}~\bibnamefont {Lon\v{c}ar}},\ }\bibfield  {title}
  {\enquote {\bibinfo {title} {Monolithic ultra-high-q lithium niobate
  microring resonator},}\ }\href@noop {} {\bibfield  {journal} {\bibinfo
  {journal} {Optica}\ }\textbf {\bibinfo {volume} {4}},\ \bibinfo {pages}
  {1536--1537} (\bibinfo {year} {2017})}\BibitemShut {NoStop}%
\bibitem [{Mia()}]{Mian}%
  \BibitemOpen
  \href@noop {} {}\bibinfo {note} {Private communcation with Mian
  Zhang}\BibitemShut {NoStop}%
\bibitem [{\citenamefont {Preskill}(2012)}]{preskill2012quantum}%
  \BibitemOpen
  \bibfield  {author} {\bibinfo {author} {\bibfnamefont {J.}~\bibnamefont
  {Preskill}},\ }\bibfield  {title} {\enquote {\bibinfo {title} {Quantum
  computing and the entanglement frontier},}\ }\href@noop {} {\bibfield
  {journal} {\bibinfo  {journal} {arXiv preprint arXiv:1203.5813}\ } (\bibinfo
  {year} {2012})}\BibitemShut {NoStop}%
\bibitem [{\citenamefont {Harrow}\ and\ \citenamefont
  {Montanaro}(2017)}]{harrow2017quantum}%
  \BibitemOpen
  \bibfield  {author} {\bibinfo {author} {\bibfnamefont {A.~W.}\ \bibnamefont
  {Harrow}}\ and\ \bibinfo {author} {\bibfnamefont {A.}~\bibnamefont
  {Montanaro}},\ }\bibfield  {title} {\enquote {\bibinfo {title} {Quantum
  computational supremacy},}\ }\href@noop {} {\bibfield  {journal} {\bibinfo
  {journal} {Nature}\ }\textbf {\bibinfo {volume} {549}},\ \bibinfo {pages}
  {203} (\bibinfo {year} {2017})}\BibitemShut {NoStop}%
\bibitem [{\citenamefont {Carolan}\ \emph {et~al.}(2015)\citenamefont
  {Carolan}, \citenamefont {Harrold}, \citenamefont {Sparrow}, \citenamefont
  {Mart{\'\i}n-L{\'o}pez}, \citenamefont {Russell}, \citenamefont
  {Silverstone}, \citenamefont {Shadbolt}, \citenamefont {Matsuda},
  \citenamefont {Oguma}, \citenamefont {Itoh}, \citenamefont {Marshall},
  \citenamefont {Thompson}, \citenamefont {Matthews}, \citenamefont
  {Hashimoto}, \citenamefont {O'Brien},\ and\ \citenamefont
  {Laing}}]{carolan2015universal}%
  \BibitemOpen
  \bibfield  {author} {\bibinfo {author} {\bibfnamefont {J.}~\bibnamefont
  {Carolan}}, \bibinfo {author} {\bibfnamefont {C.}~\bibnamefont {Harrold}},
  \bibinfo {author} {\bibfnamefont {C.}~\bibnamefont {Sparrow}}, \bibinfo
  {author} {\bibfnamefont {E.}~\bibnamefont {Mart{\'\i}n-L{\'o}pez}}, \bibinfo
  {author} {\bibfnamefont {N.~J.}\ \bibnamefont {Russell}}, \bibinfo {author}
  {\bibfnamefont {J.~W.}\ \bibnamefont {Silverstone}}, \bibinfo {author}
  {\bibfnamefont {P.~J.}\ \bibnamefont {Shadbolt}}, \bibinfo {author}
  {\bibfnamefont {N.}~\bibnamefont {Matsuda}}, \bibinfo {author} {\bibfnamefont
  {M.}~\bibnamefont {Oguma}}, \bibinfo {author} {\bibfnamefont
  {M.}~\bibnamefont {Itoh}}, \bibinfo {author} {\bibfnamefont {G.~D.}\
  \bibnamefont {Marshall}}, \bibinfo {author} {\bibfnamefont {M.~G.}\
  \bibnamefont {Thompson}}, \bibinfo {author} {\bibfnamefont {J.~C.~F.}\
  \bibnamefont {Matthews}}, \bibinfo {author} {\bibfnamefont {T.}~\bibnamefont
  {Hashimoto}}, \bibinfo {author} {\bibfnamefont {J.~L.}\ \bibnamefont
  {O'Brien}}, \ and\ \bibinfo {author} {\bibfnamefont {A.}~\bibnamefont
  {Laing}},\ }\bibfield  {title} {\enquote {\bibinfo {title} {Universal linear
  optics},}\ }\href@noop {} {\bibfield  {journal} {\bibinfo  {journal}
  {Science}\ }\textbf {\bibinfo {volume} {349}},\ \bibinfo {pages} {711--716}
  (\bibinfo {year} {2015})}\BibitemShut {NoStop}%
\bibitem [{\citenamefont {Miller}(2013)}]{miller2013self}%
  \BibitemOpen
  \bibfield  {author} {\bibinfo {author} {\bibfnamefont {D.~A.~B.}\
  \bibnamefont {Miller}},\ }\bibfield  {title} {\enquote {\bibinfo {title}
  {Self-configuring universal linear optical component},}\ }\href@noop {}
  {\bibfield  {journal} {\bibinfo  {journal} {Photon. Res.}\ }\textbf {\bibinfo
  {volume} {1}},\ \bibinfo {pages} {1--15} (\bibinfo {year}
  {2013})}\BibitemShut {NoStop}%
\bibitem [{\citenamefont {Miller}(2015)}]{miller2015perfect}%
  \BibitemOpen
  \bibfield  {author} {\bibinfo {author} {\bibfnamefont {D.~A.~B.}\
  \bibnamefont {Miller}},\ }\bibfield  {title} {\enquote {\bibinfo {title}
  {Perfect optics with imperfect components},}\ }\href@noop {} {\bibfield
  {journal} {\bibinfo  {journal} {Optica}\ }\textbf {\bibinfo {volume} {2}},\
  \bibinfo {pages} {747--750} (\bibinfo {year} {2015})}\BibitemShut {NoStop}%
\bibitem [{\citenamefont {Shen}\ \emph {et~al.}(2017)\citenamefont {Shen},
  \citenamefont {Harris}, \citenamefont {Skirlo}, \citenamefont {Prabhu},
  \citenamefont {Baehr-Jones}, \citenamefont {Hochberg}, \citenamefont {Sun},
  \citenamefont {Zhao}, \citenamefont {Larochelle}, \citenamefont {Englund},\
  and\ \citenamefont {Solja\v{c}i\'{c}}}]{shen2017deep}%
  \BibitemOpen
  \bibfield  {author} {\bibinfo {author} {\bibfnamefont {Y.}~\bibnamefont
  {Shen}}, \bibinfo {author} {\bibfnamefont {N.~C.}\ \bibnamefont {Harris}},
  \bibinfo {author} {\bibfnamefont {S.}~\bibnamefont {Skirlo}}, \bibinfo
  {author} {\bibfnamefont {M.}~\bibnamefont {Prabhu}}, \bibinfo {author}
  {\bibfnamefont {T.}~\bibnamefont {Baehr-Jones}}, \bibinfo {author}
  {\bibfnamefont {M.}~\bibnamefont {Hochberg}}, \bibinfo {author}
  {\bibfnamefont {X.}~\bibnamefont {Sun}}, \bibinfo {author} {\bibfnamefont
  {S.}~\bibnamefont {Zhao}}, \bibinfo {author} {\bibfnamefont {H.}~\bibnamefont
  {Larochelle}}, \bibinfo {author} {\bibfnamefont {D.}~\bibnamefont {Englund}},
  \ and\ \bibinfo {author} {\bibfnamefont {M.}~\bibnamefont
  {Solja\v{c}i\'{c}}},\ }\bibfield  {title} {\enquote {\bibinfo {title} {Deep
  learning with coherent nanophotonic circuits},}\ }\href@noop {} {\bibfield
  {journal} {\bibinfo  {journal} {Nat. Photonics}\ }\textbf {\bibinfo {volume}
  {11}},\ \bibinfo {pages} {441} (\bibinfo {year} {2017})}\BibitemShut
  {NoStop}%
\bibitem [{\citenamefont {Aaronson}\ and\ \citenamefont
  {Arkhipov}(2011)}]{aaronson2011computational}%
  \BibitemOpen
  \bibfield  {author} {\bibinfo {author} {\bibfnamefont {S.}~\bibnamefont
  {Aaronson}}\ and\ \bibinfo {author} {\bibfnamefont {A.}~\bibnamefont
  {Arkhipov}},\ }\bibfield  {title} {\enquote {\bibinfo {title} {The
  computational complexity of linear optics},}\ }in\ \href@noop {} {\emph
  {\bibinfo {booktitle} {Proceedings of the forty-third annual ACM symposium on
  Theory of computing}}}\ (\bibinfo {organization} {ACM},\ \bibinfo {year}
  {2011})\ pp.\ \bibinfo {pages} {333--342}\BibitemShut {NoStop}%
\bibitem [{\citenamefont {Sipser}(2006)}]{sipser2006introduction}%
  \BibitemOpen
  \bibfield  {author} {\bibinfo {author} {\bibfnamefont {M.}~\bibnamefont
  {Sipser}},\ }\href@noop {} {\emph {\bibinfo {title} {Introduction to the
  Theory of Computation}}},\ Vol.~\bibinfo {volume} {2}\ (\bibinfo  {publisher}
  {Thomson Course Technology Boston},\ \bibinfo {year} {2006})\BibitemShut
  {NoStop}%
\bibitem [{\citenamefont {Valiant}(1979)}]{valiant1979complexity}%
  \BibitemOpen
  \bibfield  {author} {\bibinfo {author} {\bibfnamefont {L.~G.}\ \bibnamefont
  {Valiant}},\ }\bibfield  {title} {\enquote {\bibinfo {title} {The complexity
  of computing the permanent},}\ }\href@noop {} {\bibfield  {journal} {\bibinfo
   {journal} {Theor. Comput. Sci.}\ }\textbf {\bibinfo {volume} {8}},\ \bibinfo
  {pages} {189--201} (\bibinfo {year} {1979})}\BibitemShut {NoStop}%
\bibitem [{\citenamefont {Rahimi-Keshari}\ \emph {et~al.}(2016)\citenamefont
  {Rahimi-Keshari}, \citenamefont {Ralph},\ and\ \citenamefont
  {Caves}}]{rahimi2016sufficient}%
  \BibitemOpen
  \bibfield  {author} {\bibinfo {author} {\bibfnamefont {S.}~\bibnamefont
  {Rahimi-Keshari}}, \bibinfo {author} {\bibfnamefont {T.~C.}\ \bibnamefont
  {Ralph}}, \ and\ \bibinfo {author} {\bibfnamefont {C.~M.}\ \bibnamefont
  {Caves}},\ }\bibfield  {title} {\enquote {\bibinfo {title} {Sufficient
  conditions for efficient classical simulation of quantum optics},}\
  }\href@noop {} {\bibfield  {journal} {\bibinfo  {journal} {Phys. Rev. X}\
  }\textbf {\bibinfo {volume} {6}},\ \bibinfo {pages} {021039} (\bibinfo {year}
  {2016})}\BibitemShut {NoStop}%
\bibitem [{\citenamefont {Oszmaniec}\ and\ \citenamefont
  {Brod}(2018)}]{oszmaniec2018classical}%
  \BibitemOpen
  \bibfield  {author} {\bibinfo {author} {\bibfnamefont {M.}~\bibnamefont
  {Oszmaniec}}\ and\ \bibinfo {author} {\bibfnamefont {D.~J.}\ \bibnamefont
  {Brod}},\ }\bibfield  {title} {\enquote {\bibinfo {title} {Classical
  simulation of photonic linear optics with lost particles},}\ }\href@noop {}
  {\bibfield  {journal} {\bibinfo  {journal} {arXiv preprint arXiv:1801.06166}\
  } (\bibinfo {year} {2018})}\BibitemShut {NoStop}%
\bibitem [{\citenamefont {Neville}\ \emph {et~al.}(2017)\citenamefont
  {Neville}, \citenamefont {Sparrow}, \citenamefont {Clifford}, \citenamefont
  {Johnston}, \citenamefont {Birchall}, \citenamefont {Montanaro},\ and\
  \citenamefont {Laing}}]{neville2017classical}%
  \BibitemOpen
  \bibfield  {author} {\bibinfo {author} {\bibfnamefont {A.}~\bibnamefont
  {Neville}}, \bibinfo {author} {\bibfnamefont {C.}~\bibnamefont {Sparrow}},
  \bibinfo {author} {\bibfnamefont {R.}~\bibnamefont {Clifford}}, \bibinfo
  {author} {\bibfnamefont {E.}~\bibnamefont {Johnston}}, \bibinfo {author}
  {\bibfnamefont {P.~M.}\ \bibnamefont {Birchall}}, \bibinfo {author}
  {\bibfnamefont {A.}~\bibnamefont {Montanaro}}, \ and\ \bibinfo {author}
  {\bibfnamefont {A.}~\bibnamefont {Laing}},\ }\bibfield  {title} {\enquote
  {\bibinfo {title} {Classical boson sampling algorithms with superior
  performance to near-term experiments},}\ }\href@noop {} {\bibfield  {journal}
  {\bibinfo  {journal} {Nat. Phys.}\ }\textbf {\bibinfo {volume} {13}},\
  \bibinfo {pages} {1153} (\bibinfo {year} {2017})}\BibitemShut {NoStop}%
\bibitem [{\citenamefont {Clifford}\ and\ \citenamefont
  {Clifford}(2018)}]{clifford2018classical}%
  \BibitemOpen
  \bibfield  {author} {\bibinfo {author} {\bibfnamefont {P.}~\bibnamefont
  {Clifford}}\ and\ \bibinfo {author} {\bibfnamefont {R.}~\bibnamefont
  {Clifford}},\ }\bibfield  {title} {\enquote {\bibinfo {title} {The classical
  complexity of boson sampling},}\ }in\ \href@noop {} {\emph {\bibinfo
  {booktitle} {Proceedings of the Twenty-Ninth Annual ACM-SIAM Symposium on
  Discrete Algorithms}}}\ (\bibinfo {organization} {Society for Industrial and
  Applied Mathematics},\ \bibinfo {year} {2018})\ pp.\ \bibinfo {pages}
  {146--155}\BibitemShut {NoStop}%
\bibitem [{\citenamefont {Lund}\ \emph {et~al.}(2014)\citenamefont {Lund},
  \citenamefont {Laing}, \citenamefont {Rahimi-Keshari}, \citenamefont
  {Rudolph}, \citenamefont {O'Brien},\ and\ \citenamefont
  {Ralph}}]{lund2014boson}%
  \BibitemOpen
  \bibfield  {author} {\bibinfo {author} {\bibfnamefont {A.~P.}\ \bibnamefont
  {Lund}}, \bibinfo {author} {\bibfnamefont {A.}~\bibnamefont {Laing}},
  \bibinfo {author} {\bibfnamefont {S.}~\bibnamefont {Rahimi-Keshari}},
  \bibinfo {author} {\bibfnamefont {T.}~\bibnamefont {Rudolph}}, \bibinfo
  {author} {\bibfnamefont {J.~L.}\ \bibnamefont {O'Brien}}, \ and\ \bibinfo
  {author} {\bibfnamefont {T.~C.}\ \bibnamefont {Ralph}},\ }\bibfield  {title}
  {\enquote {\bibinfo {title} {Boson sampling from a gaussian state},}\
  }\href@noop {} {\bibfield  {journal} {\bibinfo  {journal} {Phys. Rev. Lett.}\
  }\textbf {\bibinfo {volume} {113}},\ \bibinfo {pages} {100502} (\bibinfo
  {year} {2014})}\BibitemShut {NoStop}%
\bibitem [{\citenamefont {Hamilton}\ \emph {et~al.}(2017)\citenamefont
  {Hamilton}, \citenamefont {Kruse}, \citenamefont {Sansoni}, \citenamefont
  {Barkhofen}, \citenamefont {Silberhorn},\ and\ \citenamefont
  {Jex}}]{hamilton2017gaussian}%
  \BibitemOpen
  \bibfield  {author} {\bibinfo {author} {\bibfnamefont {C.~S.}\ \bibnamefont
  {Hamilton}}, \bibinfo {author} {\bibfnamefont {R.}~\bibnamefont {Kruse}},
  \bibinfo {author} {\bibfnamefont {L.}~\bibnamefont {Sansoni}}, \bibinfo
  {author} {\bibfnamefont {S.}~\bibnamefont {Barkhofen}}, \bibinfo {author}
  {\bibfnamefont {C.}~\bibnamefont {Silberhorn}}, \ and\ \bibinfo {author}
  {\bibfnamefont {I.}~\bibnamefont {Jex}},\ }\bibfield  {title} {\enquote
  {\bibinfo {title} {Gaussian boson sampling},}\ }\href@noop {} {\bibfield
  {journal} {\bibinfo  {journal} {Phys. Rev. Lett.}\ }\textbf {\bibinfo
  {volume} {119}},\ \bibinfo {pages} {170501} (\bibinfo {year}
  {2017})}\BibitemShut {NoStop}%
\bibitem [{\citenamefont {Kruse}\ \emph {et~al.}(2018)\citenamefont {Kruse},
  \citenamefont {Hamilton}, \citenamefont {Sansoni}, \citenamefont {Barkhofen},
  \citenamefont {Silberhorn},\ and\ \citenamefont {Jex}}]{kruse2018detailed}%
  \BibitemOpen
  \bibfield  {author} {\bibinfo {author} {\bibfnamefont {R.}~\bibnamefont
  {Kruse}}, \bibinfo {author} {\bibfnamefont {C.~S.}\ \bibnamefont {Hamilton}},
  \bibinfo {author} {\bibfnamefont {L.}~\bibnamefont {Sansoni}}, \bibinfo
  {author} {\bibfnamefont {S.}~\bibnamefont {Barkhofen}}, \bibinfo {author}
  {\bibfnamefont {C.}~\bibnamefont {Silberhorn}}, \ and\ \bibinfo {author}
  {\bibfnamefont {I.}~\bibnamefont {Jex}},\ }\bibfield  {title} {\enquote
  {\bibinfo {title} {A detailed study of {G}aussian boson sampling},}\
  }\href@noop {} {\bibfield  {journal} {\bibinfo  {journal} {arXiv preprint
  arXiv:1801.07488}\ } (\bibinfo {year} {2018})}\BibitemShut {NoStop}%
\bibitem [{\citenamefont {Barvinok}(2016)}]{barvinok2016combinatorics}%
  \BibitemOpen
  \bibfield  {author} {\bibinfo {author} {\bibfnamefont {A.}~\bibnamefont
  {Barvinok}},\ }\href@noop {} {\emph {\bibinfo {title} {Combinatorics and
  complexity of partition functions}}},\ Vol.\ \bibinfo {volume} {276}\
  (\bibinfo  {publisher} {Springer},\ \bibinfo {year} {2016})\BibitemShut
  {NoStop}%
\bibitem [{\citenamefont {Bj{\"o}rklund}\ \emph {et~al.}(2018)\citenamefont
  {Bj{\"o}rklund}, \citenamefont {Gupt},\ and\ \citenamefont
  {Quesada}}]{bjorklund2018faster}%
  \BibitemOpen
  \bibfield  {author} {\bibinfo {author} {\bibfnamefont {A.}~\bibnamefont
  {Bj{\"o}rklund}}, \bibinfo {author} {\bibfnamefont {B.}~\bibnamefont {Gupt}},
  \ and\ \bibinfo {author} {\bibfnamefont {N.}~\bibnamefont {Quesada}},\
  }\bibfield  {title} {\enquote {\bibinfo {title} {A faster hafnian formula for
  complex matrices and its benchmarking on the titan supercomputer},}\
  }\href@noop {} {\bibfield  {journal} {\bibinfo  {journal} {arXiv preprint
  arXiv:1805.12498}\ } (\bibinfo {year} {2018})}\BibitemShut {NoStop}%
\bibitem [{\citenamefont {Gupt}\ \emph {et~al.}(2018)\citenamefont {Gupt},
  \citenamefont {Arrazola}, \citenamefont {Quesada},\ and\ \citenamefont
  {Bromley}}]{gupt2018classical}%
  \BibitemOpen
  \bibfield  {author} {\bibinfo {author} {\bibfnamefont {B.}~\bibnamefont
  {Gupt}}, \bibinfo {author} {\bibfnamefont {J.~M.}\ \bibnamefont {Arrazola}},
  \bibinfo {author} {\bibfnamefont {N.}~\bibnamefont {Quesada}}, \ and\
  \bibinfo {author} {\bibfnamefont {T.~R.}\ \bibnamefont {Bromley}},\
  }\bibfield  {title} {\enquote {\bibinfo {title} {Classical benchmarking of
  {G}aussian boson sampling on the {T}itan supercomputer},}\ }\href@noop {}
  {\bibfield  {journal} {\bibinfo  {journal} {arXiv preprint arXiv:1810.00900}\
  } (\bibinfo {year} {2018})}\BibitemShut {NoStop}%
\bibitem [{\citenamefont {Arrazola}\ and\ \citenamefont
  {Bromley}(2018)}]{arrazola2018using}%
  \BibitemOpen
  \bibfield  {author} {\bibinfo {author} {\bibfnamefont {J.~M.}\ \bibnamefont
  {Arrazola}}\ and\ \bibinfo {author} {\bibfnamefont {T.~R.}\ \bibnamefont
  {Bromley}},\ }\bibfield  {title} {\enquote {\bibinfo {title} {Using
  {G}aussian boson sampling to find dense subgraphs},}\ }\href@noop {}
  {\bibfield  {journal} {\bibinfo  {journal} {arXiv preprint arXiv:1803.10730}\
  } (\bibinfo {year} {2018})}\BibitemShut {NoStop}%
\bibitem [{\citenamefont {Arrazola}\ \emph {et~al.}(2018)\citenamefont
  {Arrazola}, \citenamefont {Bromley},\ and\ \citenamefont
  {Rebentrost}}]{arrazola2018quantum}%
  \BibitemOpen
  \bibfield  {author} {\bibinfo {author} {\bibfnamefont {J.~M.}\ \bibnamefont
  {Arrazola}}, \bibinfo {author} {\bibfnamefont {T.~R.}\ \bibnamefont
  {Bromley}}, \ and\ \bibinfo {author} {\bibfnamefont {P.}~\bibnamefont
  {Rebentrost}},\ }\bibfield  {title} {\enquote {\bibinfo {title} {Quantum
  approximate optimization with {G}aussian boson sampling},}\ }\href@noop {}
  {\bibfield  {journal} {\bibinfo  {journal} {arXiv preprint arXiv:1803.10731}\
  } (\bibinfo {year} {2018})}\BibitemShut {NoStop}%
\bibitem [{\citenamefont {Br{\'a}dler}\ \emph {et~al.}(2018)\citenamefont
  {Br{\'a}dler}, \citenamefont {Dallaire-Demers}, \citenamefont {Rebentrost},
  \citenamefont {Su},\ and\ \citenamefont {Weedbrook}}]{bradler2018gaussian}%
  \BibitemOpen
  \bibfield  {author} {\bibinfo {author} {\bibfnamefont {K.}~\bibnamefont
  {Br{\'a}dler}}, \bibinfo {author} {\bibfnamefont {P.-L.}\ \bibnamefont
  {Dallaire-Demers}}, \bibinfo {author} {\bibfnamefont {P.}~\bibnamefont
  {Rebentrost}}, \bibinfo {author} {\bibfnamefont {D.}~\bibnamefont {Su}}, \
  and\ \bibinfo {author} {\bibfnamefont {C.}~\bibnamefont {Weedbrook}},\
  }\bibfield  {title} {\enquote {\bibinfo {title} {Gaussian boson sampling for
  perfect matchings of arbitrary graphs},}\ }\href@noop {} {\bibfield
  {journal} {\bibinfo  {journal} {Phys. Rev. A}\ }\textbf {\bibinfo {volume}
  {98}},\ \bibinfo {pages} {032310} (\bibinfo {year} {2018})}\BibitemShut
  {NoStop}%
\bibitem [{\citenamefont {Br\'adler}\ \emph {et~al.}(2018)\citenamefont
  {Br\'adler}, \citenamefont {Friedland}, \citenamefont {Izaac}, \citenamefont
  {Killoran},\ and\ \citenamefont {Su}}]{bradler2018graph}%
  \BibitemOpen
  \bibfield  {author} {\bibinfo {author} {\bibfnamefont {K.}~\bibnamefont
  {Br\'adler}}, \bibinfo {author} {\bibfnamefont {S.}~\bibnamefont
  {Friedland}}, \bibinfo {author} {\bibfnamefont {J.}~\bibnamefont {Izaac}},
  \bibinfo {author} {\bibfnamefont {N.}~\bibnamefont {Killoran}}, \ and\
  \bibinfo {author} {\bibfnamefont {D.}~\bibnamefont {Su}},\ }\bibfield
  {title} {\enquote {\bibinfo {title} {Graph isomorphism and {G}aussian boson
  sampling},}\ }\href@noop {} {\bibfield  {journal} {\bibinfo  {journal} {arXiv
  preprint arXiv:1810.10644}\ } (\bibinfo {year} {2018})}\BibitemShut {NoStop}%
\bibitem [{\citenamefont {Huh}\ \emph {et~al.}(2015)\citenamefont {Huh},
  \citenamefont {Guerreschi}, \citenamefont {Peropadre}, \citenamefont
  {McClean},\ and\ \citenamefont {Aspuru-Guzik}}]{huh2015boson}%
  \BibitemOpen
  \bibfield  {author} {\bibinfo {author} {\bibfnamefont {J.}~\bibnamefont
  {Huh}}, \bibinfo {author} {\bibfnamefont {G.~G.}\ \bibnamefont {Guerreschi}},
  \bibinfo {author} {\bibfnamefont {B.}~\bibnamefont {Peropadre}}, \bibinfo
  {author} {\bibfnamefont {J.~R.}\ \bibnamefont {McClean}}, \ and\ \bibinfo
  {author} {\bibfnamefont {A.}~\bibnamefont {Aspuru-Guzik}},\ }\bibfield
  {title} {\enquote {\bibinfo {title} {Boson sampling for molecular vibronic
  spectra},}\ }\href@noop {} {\bibfield  {journal} {\bibinfo  {journal} {Nat.
  Photonics}\ }\textbf {\bibinfo {volume} {9}},\ \bibinfo {pages} {615}
  (\bibinfo {year} {2015})}\BibitemShut {NoStop}%
\bibitem [{\citenamefont {Huh}\ and\ \citenamefont
  {Yung}(2017)}]{huh2017vibronic}%
  \BibitemOpen
  \bibfield  {author} {\bibinfo {author} {\bibfnamefont {J.}~\bibnamefont
  {Huh}}\ and\ \bibinfo {author} {\bibfnamefont {M.-H.}\ \bibnamefont {Yung}},\
  }\bibfield  {title} {\enquote {\bibinfo {title} {Vibronic boson sampling:
  Generalized gaussian boson sampling for molecular vibronic spectra at finite
  temperature},}\ }\href@noop {} {\bibfield  {journal} {\bibinfo  {journal}
  {Sci. Rep.}\ }\textbf {\bibinfo {volume} {7}},\ \bibinfo {pages} {7462}
  (\bibinfo {year} {2017})}\BibitemShut {NoStop}%
\bibitem [{\citenamefont {Sabapathy}\ \emph {et~al.}(2018)\citenamefont
  {Sabapathy}, \citenamefont {Qi}, \citenamefont {Izaac},\ and\ \citenamefont
  {Weedbrook}}]{sabapathy2018near}%
  \BibitemOpen
  \bibfield  {author} {\bibinfo {author} {\bibfnamefont {Krishna~Kumar}\
  \bibnamefont {Sabapathy}}, \bibinfo {author} {\bibfnamefont {Haoyu}\
  \bibnamefont {Qi}}, \bibinfo {author} {\bibfnamefont {Josh}\ \bibnamefont
  {Izaac}}, \ and\ \bibinfo {author} {\bibfnamefont {Christian}\ \bibnamefont
  {Weedbrook}},\ }\bibfield  {title} {\enquote {\bibinfo {title}
  {Near-deterministic production of universal quantum photonic gates enhanced
  by machine learning},}\ }\href@noop {} {\bibfield  {journal} {\bibinfo
  {journal} {arXiv preprint arXiv:1809.04680}\ } (\bibinfo {year}
  {2018})}\BibitemShut {NoStop}%
\bibitem [{\citenamefont {Gottesman}\ \emph {et~al.}(2001)\citenamefont
  {Gottesman}, \citenamefont {Kitaev},\ and\ \citenamefont
  {Preskill}}]{GKP2001}%
  \BibitemOpen
  \bibfield  {author} {\bibinfo {author} {\bibfnamefont {D.}~\bibnamefont
  {Gottesman}}, \bibinfo {author} {\bibfnamefont {A.}~\bibnamefont {Kitaev}}, \
  and\ \bibinfo {author} {\bibfnamefont {J.}~\bibnamefont {Preskill}},\
  }\bibfield  {title} {\enquote {\bibinfo {title} {Encoding a qubit in an
  oscillator},}\ }\href@noop {} {\bibfield  {journal} {\bibinfo  {journal}
  {Physical Review A}\ }\textbf {\bibinfo {volume} {64}},\ \bibinfo {pages}
  {012310} (\bibinfo {year} {2001})}\BibitemShut {NoStop}%
\bibitem [{\citenamefont {Marshall}\ \emph {et~al.}(2015)\citenamefont
  {Marshall}, \citenamefont {Pooser}, \citenamefont {Siopsis},\ and\
  \citenamefont {Weedbrook}}]{Marshall2015}%
  \BibitemOpen
  \bibfield  {author} {\bibinfo {author} {\bibfnamefont {K.}~\bibnamefont
  {Marshall}}, \bibinfo {author} {\bibfnamefont {M.}~\bibnamefont {Pooser}},
  \bibinfo {author} {\bibfnamefont {G.}~\bibnamefont {Siopsis}}, \ and\
  \bibinfo {author} {\bibfnamefont {C.}~\bibnamefont {Weedbrook}},\ }\bibfield
  {title} {\enquote {\bibinfo {title} {Repeat-until-success cubic phase gate
  for universal continuous-variable quantum computation},}\ }\href@noop {}
  {\bibfield  {journal} {\bibinfo  {journal} {Phys. Rev. A}\ }\textbf {\bibinfo
  {volume} {91}},\ \bibinfo {pages} {032321} (\bibinfo {year}
  {2015})}\BibitemShut {NoStop}%
\bibitem [{\citenamefont {Miyata}\ \emph {et~al.}(2016)\citenamefont {Miyata},
  \citenamefont {Ogawa}, \citenamefont {Marek}, \citenamefont {Filip},
  \citenamefont {Yonezawa}, \citenamefont {Yoshikawa},\ and\ \citenamefont
  {Furusawa}}]{MiyataAdapNonGauss2016}%
  \BibitemOpen
  \bibfield  {author} {\bibinfo {author} {\bibfnamefont {K.}~\bibnamefont
  {Miyata}}, \bibinfo {author} {\bibfnamefont {H.}~\bibnamefont {Ogawa}},
  \bibinfo {author} {\bibfnamefont {P.}~\bibnamefont {Marek}}, \bibinfo
  {author} {\bibfnamefont {R.}~\bibnamefont {Filip}}, \bibinfo {author}
  {\bibfnamefont {H.}~\bibnamefont {Yonezawa}}, \bibinfo {author}
  {\bibfnamefont {J.}~\bibnamefont {Yoshikawa}}, \ and\ \bibinfo {author}
  {\bibfnamefont {A.}~\bibnamefont {Furusawa}},\ }\bibfield  {title} {\enquote
  {\bibinfo {title} {Implementation of a quantum cubic gate by an adaptive
  non-{G}aussian measurement},}\ }\href@noop {} {\bibfield  {journal} {\bibinfo
   {journal} {Phys. Rev. A}\ }\textbf {\bibinfo {volume} {93}},\ \bibinfo
  {pages} {022301} (\bibinfo {year} {2016})}\BibitemShut {NoStop}%
\bibitem [{\citenamefont {Arzani}\ \emph {et~al.}(2017)\citenamefont {Arzani},
  \citenamefont {Treps},\ and\ \citenamefont {Ferrini}}]{Arzani2017}%
  \BibitemOpen
  \bibfield  {author} {\bibinfo {author} {\bibfnamefont {F.}~\bibnamefont
  {Arzani}}, \bibinfo {author} {\bibfnamefont {N.}~\bibnamefont {Treps}}, \
  and\ \bibinfo {author} {\bibfnamefont {G.}~\bibnamefont {Ferrini}},\
  }\bibfield  {title} {\enquote {\bibinfo {title} {Polynomial approximation of
  non-{G}aussian unitaries by counting one photon at a time},}\ }\href@noop {}
  {\bibfield  {journal} {\bibinfo  {journal} {Phys. Rev. A}\ }\textbf {\bibinfo
  {volume} {95}},\ \bibinfo {pages} {052352} (\bibinfo {year}
  {2017})}\BibitemShut {NoStop}%
\bibitem [{\citenamefont {Nielsen}(2018)}]{nielsen2018neural}%
  \BibitemOpen
  \bibfield  {author} {\bibinfo {author} {\bibfnamefont {M.~A.}\ \bibnamefont
  {Nielsen}},\ }\href {http://neuralnetworksanddeeplearning.com/} {\enquote
  {\bibinfo {title} {Neural networks and deep learning},}\ } (\bibinfo {year}
  {2018})\BibitemShut {NoStop}%
\bibitem [{\citenamefont {Yeh}\ \emph {et~al.}(2004)\citenamefont {Yeh},
  \citenamefont {Lo},\ and\ \citenamefont {Shi}}]{yeh2004optical}%
  \BibitemOpen
  \bibfield  {author} {\bibinfo {author} {\bibfnamefont {S.~L.}\ \bibnamefont
  {Yeh}}, \bibinfo {author} {\bibfnamefont {R.~C.}\ \bibnamefont {Lo}}, \ and\
  \bibinfo {author} {\bibfnamefont {C.~Y.}\ \bibnamefont {Shi}},\ }\bibfield
  {title} {\enquote {\bibinfo {title} {Optical implementation of the hopfield
  neural network with matrix gratings},}\ }\href@noop {} {\bibfield  {journal}
  {\bibinfo  {journal} {Appl. Opt.}\ }\textbf {\bibinfo {volume} {43}},\
  \bibinfo {pages} {858--865} (\bibinfo {year} {2004})}\BibitemShut {NoStop}%
\end{thebibliography}%

\end{document}